\documentclass[fleqn,usenatbib]{mnras}

\usepackage{newtxtext,newtxmath}

\usepackage{orcidlink}
\usepackage{balance}

\usepackage[T1]{fontenc}

\DeclareRobustCommand{\VAN}[3]{#2}
\let\VANthebibliography\thebibliography
\def\thebibliography{\DeclareRobustCommand{\VAN}[3]{##3}\VANthebibliography}

\usepackage{graphicx}	
\usepackage{amsmath}	

\title[ACF of Lyman-break galaxies at $3 \lesssim z \lesssim 5$]{Angular correlation functions of bright Lyman-break galaxies at $\mathbf{3 \lesssim z \lesssim 5}$}

\author[Isabelle Ye et al.]{Isabelle Ye\,\orcidlink{0009-0007-1958-3364},$^{1}$\thanks{E-mail: isabelle.ye@postgrad.manchester.ac.uk}\
Philip Bull\,\orcidlink{0000-0001-5668-3101},$^{1,2}$
Rebecca A. A. Bowler\,\orcidlink{0000-0003-3917-1678},$^{1}$
Rachel K. Cochrane\,\orcidlink{0000-0001-8855-6107},$^{1}$
Nathan J. Adams\,\orcidlink{0000-0003-4875-6272},$^{1}$
\newauthor
and Matt J. Jarvis\,\orcidlink{0000-0001-7039-9078}$^{3,2}$
\\
$^{1}$Jodrell Bank Centre for Astrophysics, University of Manchester, Manchester M13 9PL, UK\\
$^{2}$Department of Physics and Astronomy, University of Western Cape, Cape Town 7535, South Africa\\
$^{3}$Astrophysics, Department of Physics, University of Oxford, Keble Road, Oxford, OX1 3RH, UK
}

\date{Accepted XXX. Received YYY; in original form ZZZ}

\pubyear{2025}

\begin{document}
\label{firstpage}
\pagerange{\pageref{firstpage}--\pageref{lastpage}}
\maketitle

\begin{abstract}   
We investigate the clustering of Lyman-break galaxies at redshifts of 3 $\lesssim z \lesssim$ 5 within the COSMOS field by measuring the angular two-point correlation function. Our robust sample of $\sim$60,000 bright ($m_{\rm UV}\lesssim 27$) Lyman-break galaxies was selected based on spectral energy distribution fitting across 14 photometric bands spanning optical and near-infrared wavelengths. We constrained both the 1- and 2-halo terms at separations up to 300 arcsec, finding an excess in the correlation function at scales corresponding to $<20$ kpc, consistent with enhancement due to clumps in the same galaxy or interactions on this scale.
We then performed Bayesian model fits on the correlation functions to infer the Halo Occupation Distribution parameters, star formation duty cycle, and galaxy bias in three redshift bins. 
We examined several cases where different combinations of parameters were varied, showing that our data can constrain the slope of the satellite occupation function, which previous studies have fixed. For an $M_{\rm{UV}}$-limited sub-sample, we found galaxy bias values of $b_g=3.18^{+0.14}_{-0.14}$ at $z\simeq3$, $b_g=3.58^{+0.27}_{-0.29}$ at $z\simeq4$, $b_g=4.27^{+0.25}_{-0.26}$ at $z\simeq5$. The duty cycle values are $0.62^{+0.25}_{-0.26}$, $0.40^{+0.34}_{-0.22}$, and $0.39^{+0.31}_{-0.20}$,	respectively. These results suggest that, as the redshift increases, there is a slight decrease in the host halo masses and a shorter timescale for star formation in bright galaxies, at a fixed rest-frame UV luminosity threshold.

\end{abstract}




\section{Introduction}
\label{sec:introduction}

The redshift range from $z=3-5$ represents a period of rapid fractional increase in the volume-averaged star formation rate density \citep{Madau_2014}, and is therefore a crucial epoch for understanding galaxy evolution. Observing galaxies during this epoch can be challenging as the rest-frame UV light from the galaxies is redshifted into the red optical, making NIR bands increasingly important for galaxy selection. High-redshift studies to date have covered survey areas of $\sim1-30$ square degrees; examples include the JWST Advanced Deep Extragalactic Survey (JADES; \citealt{Eisenstein_2023}), the COSMOS field \citep{Weaver_2022,Shuntov_2025_2}, the the CFHT large area $U$-band deep survey (CLAUDS;\citealt{Sawicki_2019}) and the Hyper-Suprime Cam Subaru Strategic Program (HSC-SSP) survey \citep{Aihara_2018}.
On the other hand, the largest cosmological galaxy surveys such as the Dark Energy Survey (DES) and Dark Energy Spectroscopic Instrument (DESI), covering $\sim10,000$ deg$^2$, have primarily focused on $z \lesssim 2$ \citep[e.g.][]{DES_2016, DESI_2022}. With upcoming wide-area survey telescopes like $\emph{Euclid}$ and Rubin/LSST, detailed studies of galaxy populations and their environments are now becoming feasible at higher redshifts \citep{Ivezi_2019,Euclid_2025}.

Two-point correlation functions are a fundamental statistical tool in large-scale structure cosmology for measuring the clustering of galaxies \citep[e.g.][]{Peebles_1974, Peebles_1983,1985Natur.314..718S,1992ApJ...392..419B,  Wang_2013}. They quantify how galaxies are distributed across different scales, and can be linked to the distribution and properties of host dark matter haloes where the galaxies likely reside \citep[e.g.][]{1978MNRAS.183..341W, Guo_2010, Wechsler_2018}. They are typically denoted by $\xi(r)$ or $w(\theta)$, which describe the probability of finding a pair of galaxies at a comoving separation of $r$ or angular separation $\theta$.

On large scales, perturbations in the matter density are approximately linear: $|\delta| \ll 1$, where the density contrast $\delta$ describes the fractional excess of matter density in a given region relative to the mean density of the Universe. In this regime, linear perturbation theory can be used to accurately describe the evolution and statistics of the density field. Physical features such as the Baryon Acoustic Oscillations (e.g. \citealt{Eisenstein_1998,Seo_2003}), which can be used to reconstruct the cosmic expansion history $H(z)$ \citep{Linder_2005}, can be observed in the clustering of galaxies \citep[e.g.][]{Chuang_2013,Alcaniz:2016ryy}.

On smaller scales, in the non-linear regime ($\delta \geq 1$), typically on the order of a few Mpc \citep[$\sim$10 Mpc at $z\simeq 0$;][]{Tassev_2012}, linear perturbation theory breaks down however. Non-linear gravitational collapse leads to the formation of virialised dark matter haloes \citep[e.g.][]{Mota_2004,PhysRevD.90.083518}. The distribution of (invisible) dark matter haloes and the observed galaxy distribution can be linked through a `halo model', which predicts how galaxies of different types are expected to populate haloes as a function of halo mass. The key components of a halo model are the halo mass function (the number density of haloes as a function of mass), the halo bias (how haloes are clustered relative to the overall matter distribution), and the dark matter density profile (the internal matter distribution within a halo). Based on models of how dark matter is distributed inside haloes, the halo model can also incorporate a model of the galaxy `halo occupation distribution' (HOD;  \citealt{cooray_2002}), which describes the probability distribution of the number of galaxies residing in a dark matter halo as a function of halo mass (e.g. \citealt{Asgari_2023}).

By fitting a halo model to the observed galaxy correlation function, we can infer how galaxies are spatially distributed within dark matter haloes. The model can be split into two main contributions. The 2-halo term models correlations between different haloes and concerns larger scales, allowing us to measure the galaxy bias \citep[e.g.][]{2000MNRAS.318.1144P,Desjacques_2018}, which quantifies how galaxies cluster relative to the underlying dark matter distribution. The bias is defined on linear scales as $b_g=\delta_g/\delta$, where $\delta_g$ is the galaxy overdensity, $\delta$ is the dark matter overdensity, and $b_g$ is the galaxy bias. On smaller scales, the 1-halo term probes the internal conditions of individual haloes, particularly how different types of galaxies are distributed within them, allowing insight into galaxy formation and feedback processes \citep[e.g.][]{Hatfield_2016,Hatfield_2017}. The clustering model can also be used to constrain quantities such as the stellar-to-halo mass relation \citep[SHMR, e.g.][]{Leauthaud_2011,Wechsler_2018}.

Several studies have measured the clustering of Lyman-break galaxies (LBGs) at $3 < z < 5$ from different datasets. LBGs are effective tracers for identifying large numbers of galaxies at high redshifts due to their characteristic spectral drop-off at the Lyman limit, which allows for their detection in photometric surveys.
Early work made simple estimates of the correlation length $r_0$ by fitting the spatial two-point correlation function $\xi(r)$ with a power-law form $\xi(r) = ( {r_0}/{r} )^{\gamma}$. \cite{Ouchi_2004} studied LBGs in the Subaru Deep Field and the Subaru \textit{XMM-Newton} Deep Field and found correlation lengths of $r_0 \sim 4 - 6$ $h^{-1}$Mpc for LBGs at $z \sim 4 - 5$, and showed that brighter galaxies exhibit stronger clustering. \cite{Kashikawa_2005} also measured the clustering of LBGs at $z\simeq 4 - 5$ in the Subaru Deep Field, obtaining angular correlation functions $w(\theta)$ of comparable amplitude to these earlier results. They found large bias factors consistent with LBGs residing in dark matter haloes of $\sim 10^{11} - 10^{12}$ $M_\odot$.

Similar trends were confirmed by \cite{Lee_2006} using the GOODS fields, and by \cite{Hildebrandt_2009} with the CFHTLS-Deep survey ($\sim4$ deg$^2$), who also employed careful colour selection to limit low-$z$ interloper contamination. They found that both correlation length and inferred halo mass decrease with decreasing rest-frame UV luminosity. Likewise, \cite{Cooke_2013} using the CFHTLS Deep Fields, and \cite{Bian_2013} studying LBGs selected in the NOAO Bo\"otes field, both measured the the angular correlation function (ACF) at $z=3$. Both reported correlation lengths of $r_0 \sim 4 - 6$ $h^{-1}$Mpc, comparable with the with earlier measurements by \cite{Ouchi_2004}, and reported the same luminosity-dependent clustering trends using power-law fits to their data.

Larger spectroscopic samples, such as the VLT VIMOS LBG Survey \citep{Bielby_2011}  and the VIMOS Ultra Deep Survey \citep{Durkalec_2015}, reported $r_0 \sim 3 - 4$ $h^{-1}$ Mpc at $z \sim 3$ and inferred a galaxy bias of $b\simeq 2 - 3$. \cite{Durkalec_2015} estimated halo masses of a few times $10^{11}$ M$\odot$. More recent studies, including \cite{Harikane_2016,Harikane_2022} and \cite{Ishikawa_2017}, used large optical surveys from HSC and employed HOD models to constrain parameters such as the minimum halo mass $M_{\text{min}}$ required to host a central galaxy, finding that typical LBGs reside in haloes of $\sim 10^{11} - 10^{12}$ $M_\odot$. While clustering measurements have now extended to $z \gtrsim 6$ (e.g. \citealt{Hatfield_2018}, \citealt{Dalmasso_2024}, \citealt{Paquereau_2025}, \citealt{Shuntov_2025}), current samples at these redshifts are generally too small for robust HOD modelling.

The ACF studies at $z=3-5$ to date have relied on samples extracted from optical photometry (or in some cases spectroscopic follow-up of small samples), typically selected via colour-colour cuts to identify the Lyman break. 
However in the search for $z=4 - 5$ sources in optical imaging, there is a smaller number of detection bands into the red optical (e.g. only $i,z$ and $y$ available at $z=5$) making samples more vulnerable to contamination (e.g. from brown dwarfs; \citealp{Bowler_2015}). Furthermore, simple colour-cuts have been shown to be incomplete \citep{adams2023total} in comparison to photometric fitting that includes additional filters. 
In particular the addition of NIR filters in the selection process has many benefits at $z = 3$--$5$, providing a reduction in contamination, and the ability to derive physical properties like stellar masses from rest-frame optical imaging.

Future surveys like the Legacy Survey of Space and Time \citep[LSST;][]{Ivezi_2019} will cover large areas of the sky at $z=3-5$. The LSST's wide-deep-fast survey will observe in the $ugrizy$ bands, reaching a 5-$\sigma$ depth of 24.0 in the $r$ band with a single exposure. The coadded $r$-band image depth will reach 26.9 after 10 years, over a vast area of approximately 18,000 square degrees \citep{LSSTDarkEnergyScience:2018jkl}. This final coadded depth is comparable to the survey data we use here, but achieved across a significantly larger area.

In this work we measure ACF in the Cosmic Evolution Survey field \citep[COSMOS;][]{2007ApJS..172....1S}. Our catalogue is a sample of high redshift galaxies covering the redshift range $z$ $= 2.75 - 5.2$. It includes multi-wavelength photometry across 14 bands (optical and near-infrared) with robust selection criteria to minimise contamination. The photometric redshifts of the galaxies are obtained through SED template fitting by \cite{adams2023total}. We apply the halo model with a HOD parametrisation to fit the observed angular correlation functions, estimating the HOD parameters such as halo mass, and fitting for the galaxy bias using a Bayesian Markov Chain Monte Carlo (MCMC) approach. We also compare different sets of free and fixed parameters within the halo model to see how each parameter affects the shape of the correlation function.

The paper is structured as follows. Section~\ref{sec:galaxy_data} describes the data we used and how we sample the redshift bins. Section~\ref{sec:ACF} describes the methodology used to measure the angular correlation function and presents the main results. Section~\ref{sec:halo_model_fitting} presents the results of halo model fits to the observed ACF. We conclude in Section~\ref{sec:discussion}. We adopt the \cite{2020A&A...641A...6P} values for the cosmological parameters: $\Omega_c=0.261$, $\Omega_b=0.0490$, $h=0.677$, $\sigma_8=0.811$, $n_s=0.965$, $\rm N_{eff}=3.046$.

\section{Data and sample}
\label{sec:galaxy_data}

In this section we describe the COSMOS field dataset, including the available multi-wavelength imaging, photometric redshift estimation methods, and criteria for galaxy selection across the three redshift ranges used. We then present the specifics of the dataset, such as filter depths and galaxy counts, along with the redshift distribution of selected galaxies.

\subsection{The COSMOS field}

The COSMOS field data used in this study is a multi-wavelength dataset that spans optical and near-infrared photometry. The sample we begin with is identical to the dataset used in \cite{adams2023total}. Photometric observations include up to 14 bands, spanning effective wavelengths from 0.375 to 2.19 $\mu$m. This deep imaging enables the detection of faint galaxies with absolute UV magnitudes down to $M_{\rm{UV}} \approx -20.5$ at redshifts around $z = 5$. The field covers an area of 1.51 deg$^2$ and includes observations from several surveys. Optical data comes from the HyperSuprimeCam Strategic Survey Programme \citep[HSC DR2;][]{Aihara_2017,Aihara_2019}, and also the Canada-France-Hawaii-Telescope Legacy Survey \citep[CFHTLS;][]{2012SPIE.8448E..0MC}. The near-infrared data are sourced from the UltraVISTA survey \citep[UltraVISTA DR4;][]{McCracken_2012}. Furthermore, the overlapping datasets from HSC and UltraVISTA are leveraged for accurate photometric redshift estimates obtained through a spectral energy distribution (SED) fitting process, which we describe in Section~\ref{subsec:sample_tomobin}. Table~\ref{tab:filters_depths} provides the filters used for the selection of galaxies in \cite{adams2023total}, their 5$\sigma$ depths in AB magnitudes, and the corresponding data sources (CFHT, HSC, and VISTA).

\begin{table}
    \centering
    \begin{tabular}{|l|c|c|}
        \hline
        Filter & Depth (5$\sigma$, AB mag) & Origin \\
        \hline
        $u^*$ & 27.1 & CFHT \\
        $g^*$ & 27.3 & CFHT \\
        $r^*$ & 26.9 & CFHT \\
        $i^*$ & 26.6 & CFHT \\
        $z^*$ & 25.5 & CFHT \\
        \hline
        $g$ & 27.4 & HSC \\
        $r$ & 27.1 & HSC \\
        $i$ & 26.9 & HSC \\
        $z$ & 26.5 & HSC \\
        $y$ & 25.7 & HSC \\
        \hline
        $Y$ & 25.4 & VISTA \\
        $J$ & 25.3 & VISTA \\
        $H$ & 25.1 & VISTA \\
        $K_s$ & 25.0 & VISTA \\
        \hline
    \end{tabular}
    \caption{Filters and 5$\sigma$ depths (AB magnitudes) for the COSMOS field. These filters span from optical to near-infrared wavelengths, utilizing data from CFHT, HSC, and VISTA, which enables photometric measurements across a wide range of effective wavelengths (0.375–2.19 $\mu$m). The depths are calculated using circular $2^{\prime\prime}$ diameter apertures.}
    \label{tab:filters_depths}
\end{table}

\subsection{Sample and tomographic bin selection}
\label{subsec:sample_tomobin}

Galaxies were selected using SED fitting with \textsc{LePhare} \citep{Arnouts_1999,Ilbert_2006}, a template-fitting code that minimises $\chi^2$ values across various SED templates to compute best-fit photometric redshifts and physical parameters. For galaxies in the redshift ranges $2.75 < z < 3.5$ and $3.5 < z < 4.5$, sources were selected with a $\geq 5\sigma$ detection in the band corresponding to the rest-frame ultraviolet continuum emission (HSC $r$-band for $z \simeq 3$ and $i$-band for $z \simeq 4$). These sources were required to have a best-fitting SED template corresponding to a galaxy or QSO within the respective redshift ranges, and have a maximum $\chi^2$ value of 100 to filter out potential artifacts. For $4.5 < z < 5.2$, the initial sample consisted of objects with a best-fitting SED template corresponding to a galaxy or QSO, and have a 5$\sigma$ detection in the HSC-$z$ band. Additionally, a $<3\sigma$ detection in the CFHT-$u^*$ band was required to minimise lower redshift contaminants.

Photometry was extracted using \textsc{SExtractor} \citep{Bertin_1996} in dual image mode. For each subfield, the deepest $r, i, z$ bands are used to capture the rest-frame ultraviolet emission for $z\simeq$ 3, 4 and 5 respectively. Details of the catalogue creation can be found in \cite{Bowler_2020}. The broad wavelength range of the dataset and depth in NIR ensures robust photometric redshift estimates. The uncertainties in photometric redshifts of galaxies are described in detail in \cite{adams2023total}. They reported the normalised median absolute deviation (NMAD) of the photometric redshifts compared to a compilation of spectroscopic redshifts to be 0.029, with an outlier rate of 3.1\% for the COSMOS field. Even at the faint limit of our sample ($M_{\rm{UV}}\simeq-20$ at $z\simeq5$), sources have detected breaks of $\gtrsim1.5$ mag, which can be safely considered a robust Lyman break detection and suggests that contamination from low-redshift interlopers should be minimal.

\begin{figure}
    \centering
    \includegraphics[width=0.47\textwidth]{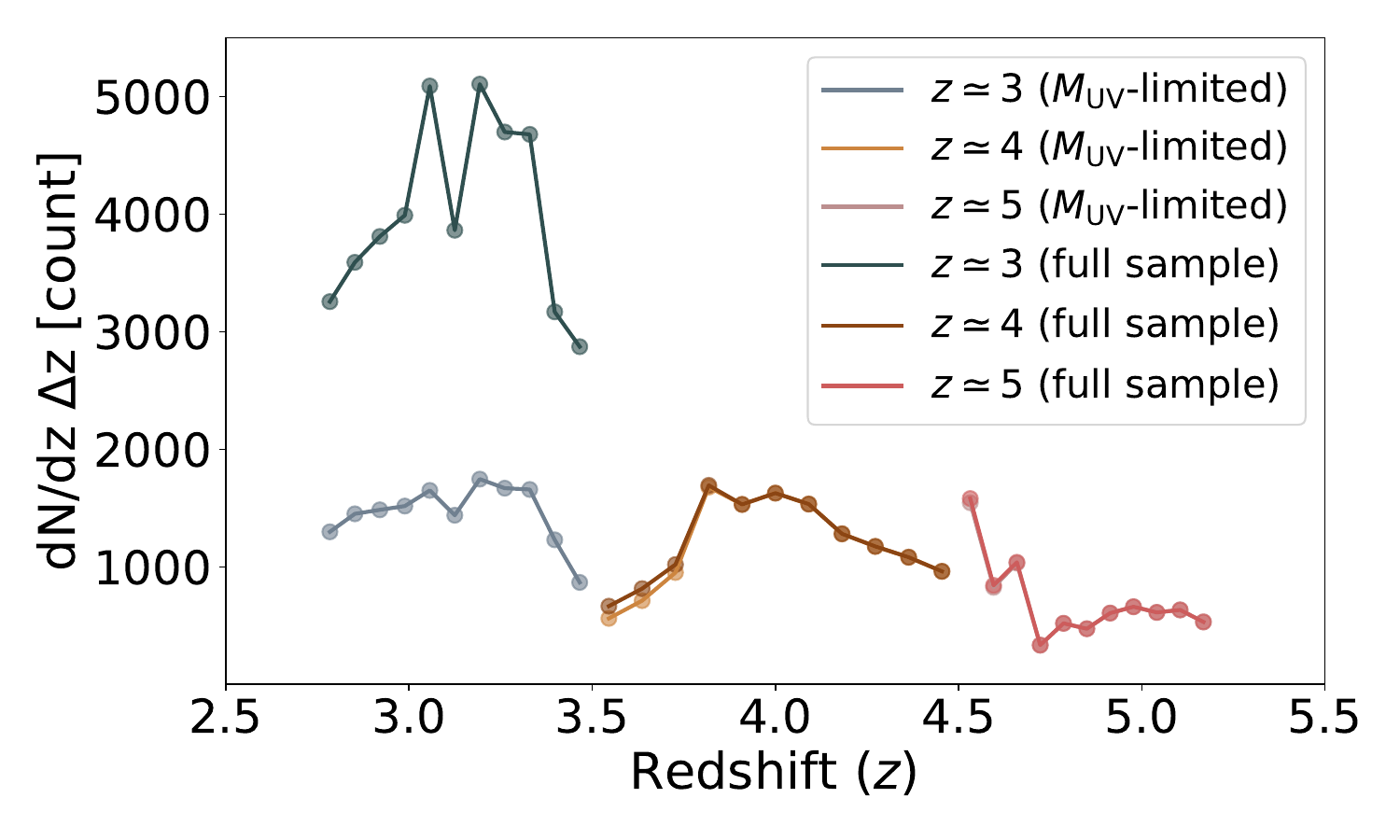}
    \caption{The redshift distribution of galaxies in the COSMOS field across our three photometric $z$ bins selected in \protect\cite{adams2023total}. The total galaxy counts for the full samples are 44,132 ($z\simeq 3$), 13,407 ($z\simeq 4$), and 7,861 ($z\simeq 5$). The galaxy counts for the $M_{\rm{UV}}$-limited samples are 16,033 ($z\simeq 3$), 13,118 ($z\simeq 4$), and 7,800 ($z\simeq 5$).}
    \label{fig:dNdz_COSMOS}
\end{figure}

The $2.5 < z < 2.75$ galaxies were excluded due to the issues related to the visibility of the Lyman break in this redshift range with the available filters \citep{adams2023total}. Additionally, the shallow $u^*$-band resulted in degeneracies with lower redshift ($z<0.4$) Balmer break features, increasing contamination of other sources. The sample also excludes data at redshifts $z > 5.2$ due to the difficulty in separating galaxies from Milky Way brown dwarfs, particularly M-class dwarfs, whose colours in the optical and near-infrared bands become comparable with those of redshift $z > 5.2$ galaxies \citep[e.g.][]{Stern_2007,Bowler_2015,Vledder_2016}. Fig.~\ref{fig:dNdz_COSMOS} shows the number of galaxies in our sample as a function of redshift for the COSMOS field across the three redshift bins.

\section{Angular correlation function}
\label{sec:ACF}

In this section, we review the angular two-point correlation function and the estimators we use to measure it and the associated measurement errors (covariance matrices). We then present the results of the two-point correlation functions measured from the LBG sample, including the covariance matrices for the three redshift bins that are needed for the HOD model inference discussed in Section \ref{sec:halo_model_fitting}.

\subsection{Correlation function estimator}

The galaxy overdensity in a pixel $i$ of a redshift bin is defined as
\begin{equation}
    \delta_g^{(i)} = \frac{n_g^{(i)}-\bar{n}_g}{\bar{n}_g},
	\label{eq:overdensity}
\end{equation}
where $n_g^{(i)}$ is the observed number density of galaxies in the pixel, and $\bar{n}_g$ is the mean number density across the redshift bin. The angular correlation function $w(|\vec{\theta}|)$ quantifies the excess probability of finding a pair of galaxies with angular separation $\theta$ compared with an uncorrelated Poisson random field. It is defined as
\begin{equation}
    w(|\Vec{\theta}|) = \langle \delta_g(\Vec{\theta} + \Vec{\theta_c}) \delta_g(\Vec{\theta_c}) \rangle.
\end{equation}
The angle brackets $\langle \ldots \rangle $ denote an average over all pairs of galaxies separated by $|\Vec{\theta}|$, where this quantity is independent of the reference position $\Vec{\theta_c}$ under the assumption of statistical isotropy and homogeneity. The average can be calculated by performing pair counts in discrete angular bins,
\begin{equation}
    w(\theta_\alpha) = \frac{\sum_{\textit{p},\textit{q}} \delta_g^\textit{(p)} \delta_g^\textit{(q)} \Theta^\alpha_{\textit{p}\textit{q}}}{\sum_{\textit{p},\textit{q}} \Theta^\alpha_{\textit{p}\textit{q}}},
    \label{eq:peebles_estimater}
\end{equation}
where the summation is performed over all unique pairs of sources labelled by $\textit{p}$ and $\textit{q}$. $\Theta^\alpha_{\textit{p}\textit{q}}$ is a function that equals 1 if the angular separation between the sources is within a angular bin labelled by $\alpha$, and 0 otherwise. In what follows, we used 18 bins, with separation values $\theta$ ranging from 1 to $10^3$~arcsec on a logarithmic scale. The bins are mostly evenly spaced, with deviations from perfectly uniform (logarithmic) spacing mostly within 2$\%$ and often below 1$\%$ (the bin edges are chosen by the {\tt TreeCorr} code; see below). For reference, an angular separation of $1^\circ$ corresponds to a comoving separation of 114~Mpc at $z\simeq3$, 128~Mpc at $z\simeq4$, and 139~Mpc at $z\simeq5$, assuming our fiducial Planck cosmology.

Survey data typically contain a number of imperfections, such as missing regions due to masking (e.g. around bright stars) and variable depth due to differing integration times, seeing conditions, and survey irregularities. In practice, $w(\theta)$ can be measured using a statistical estimator that takes into account these imperfections by referencing the measured galaxy pair counts to the pair counts for a catalogue of randomly generated positions with the same masking/depth pattern. In this work, we use the Landy–Szalay estimator \citep{1993ApJ...412...64L},
\begin{equation}
w(\theta) = \frac{N_{dd} - 2N_{dr} + N_{rr}}{N_{rr}},
\label{eq:landy_szalay_estimater}
\end{equation}
where $N_{dd}$ is the normalised number of pairs of galaxies in the observed data with an angular separation $\theta$, $N_{rr}$ is the same quantity but for the catalogue of randomly generated positions, and $N_{dr}$ is the normalised number of galaxy--random pairs, i.e. a cross-correlation between the two catalogues. Note that the two catalogues do not need to have the same number of entries; in fact, it is possible to reduce the variance of the estimator by using a random catalogue with several times more entries than the observed catalogue. This estimator is widely used due to its reduced bias and variance compared to simpler alternatives and has been shown to be better at handling edge effects compared to the Peebles estimator \citep{Peebles_1974} (e.g., \citealt{1993ApJ...412...64L,2013A&A...554A.131V,2000ApJ...535L..13K,2022A&A...666A.181K}).

For the random catalogue, we generate mock galaxies that are Poisson distributed across the survey footprint. This results in a random distribution of galaxies that are not clustered but do follow the survey geometry, i.e. we exclude any mock galaxies that fall within the masked areas of the data. This provides a reference to handle the effects of masking and the non-uniform survey geometry, although in our case, we note that the COSMOS field is relatively uniform in depth. We set the number of random galaxies to be ten times the number of observed galaxies, following \cite{Wang_2013} and \cite{10.1093/mnras/stt1933}.

\subsection{Covariance of the ACF}
\label{subsec:estimate_error}

In order to accurately estimate the halo model parameters, we also need to estimate the covariance matrix, $C$, of the binned angular correlation function. There are various approaches for this, including using large ensembles of simulations, or empirical methods like bootstrap and jackknife.

Jackknife and bootstrap error estimates account for fluctuations such as shot noise -- the random fluctuations in galaxy number density due to the discrete and finite size of the sample. Since these methods estimate the variance in galaxy number density, the shot noise is then naturally included. However, these methods do not include other sources of variance, such as super-sample covariance (SSC; \citealt{Bayer_2023}) and cosmic variance. SSC comes from the large-scale density fluctuations in the Universe. As we are sampling only a finite region of the Universe, the measured two-point function is affected by modes that are associated with fluctuations on scales larger than the survey volume. Future experiments covering much larger areas are expected to mitigate this issue. The cosmic variance arises because we can observe only a single and finite region of the Universe, which may not fully represent the cosmic average. In our analysis, we use only the jackknife method to estimate covariance, and do not include these other sources of variance in our estimates.

The jackknife method estimates the covariance matrix $C$ by breaking up the survey area into multiple patches, then excluding one patch at a time from the input catalogue, computing the correlation function for the galaxies across all the remaining patches, and finally using the variation across these estimates to calculate $C$:
\begin{equation}
    C = \frac{N_\text{patch}-1}{N_\text{patch}}\sum_{i}(w_{i}-\bar{w})^{\top}(w_{i}-\bar{w}),
	\label{eq:jackknife}
\end{equation}
where $N_{\text{patch}}$ is the total number of patches, and $w_{i}$ is the correlation function estimate with the $i$-th patch excluded. $\bar{w}$ is the mean correlation function over all patches.

For comparison, we also used the bootstrap method, which estimates the covariance matrix by resampling the data with replacement. The process involves randomly selecting $N_{\text{patch}}$ patches from the full list $[0, \ldots, N_{\text{patch}} - 1]$ with replacement. The correlation function is then computed for all galaxies in the selected patches. This resampling is repeated $N_{\text{bootstrap}}$ times, generating a set of resampled correlation functions, $\{w_j\}$. The covariance matrix is estimated as the sample variance of these resampled results:
\begin{equation}
    C = \frac{1}{N_\text{bootstrap}-1}\sum_{i}(w_{j}-\bar{w})^{\top}(w_{j}-\bar{w}).
	\label{eq:bootstrap}
\end{equation}

When selecting $N_{\text{patch}}$, the patch size should be larger than the maximum scale of the correlation function being measured. This condition suggests $N_{\text{patch}}\approx20$ for our maximum separation of 1000~arcsec, but such a choice results in poorly structured (noisy) correlation matrices for both estimators. From the clustering measurement, the last three data points have larger error bars and low signal-to-noise, and have little impact on the model fits we perform below, and so we focus on the points below 300~arcsec. This translates to dividing the full survey field into $\approx$ 300 patches. We verified that covariance matrices computed with $N_{\text{patch}}=$ 200, 300, and 400 yield consistent results, with variations of about $7\%$. Therefore, we opted to use $N_{\text{patch}}=300$ to construct the covariance matrices for the redshift bins $z\simeq3$, $z\simeq4$, and $z\simeq5$.

\begin{figure*}
    \centering
    \includegraphics[width=\textwidth]{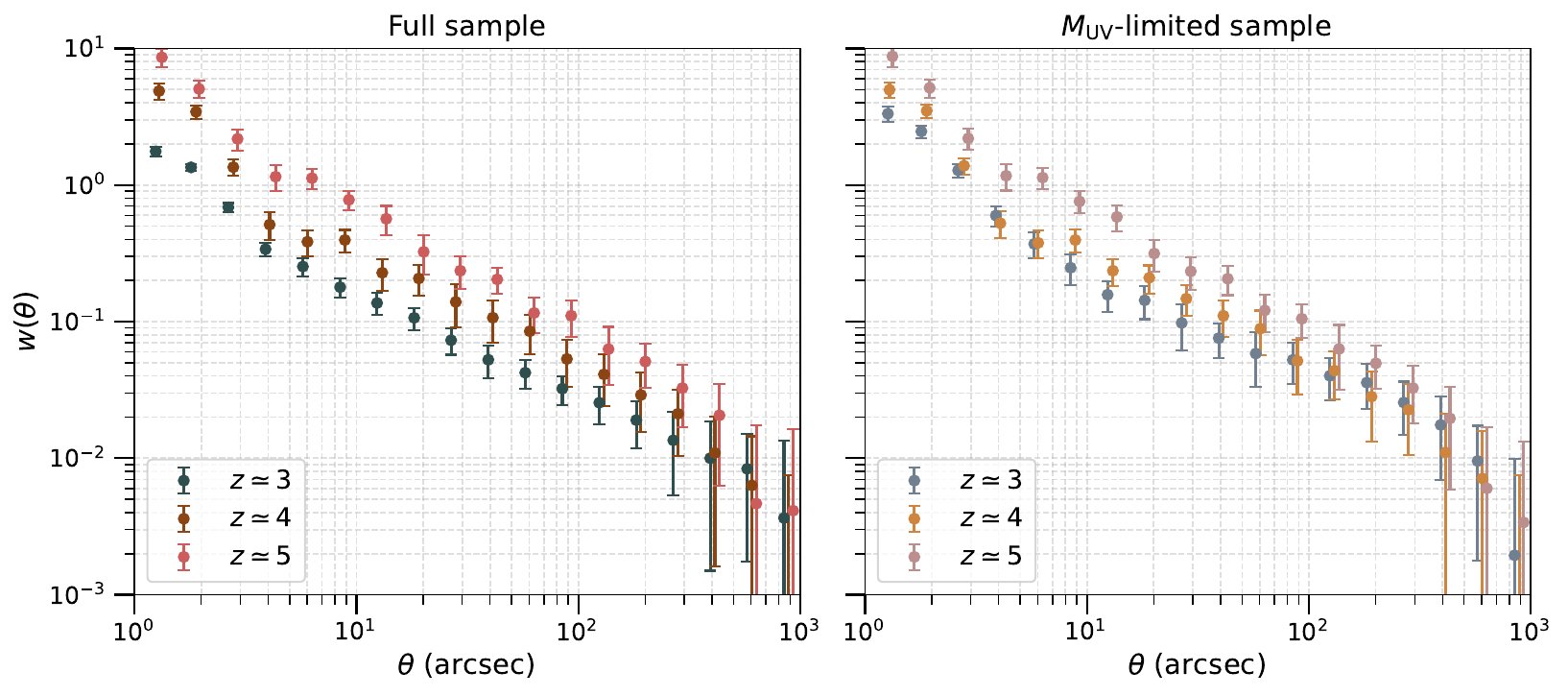}
    \caption{Angular correlation function for all galaxies in our sample of LBGs (left) and with an absolute magnitude cut of $M_{\rm{UV}}\le-20.0$ (right), shown at three redshift bins. The data points for higher redshifts are shifted slightly to the right for clarity.}
    \label{fig:w_all}
\end{figure*} 

\begin{figure*}
    \centering
    \includegraphics[width=0.98\textwidth]{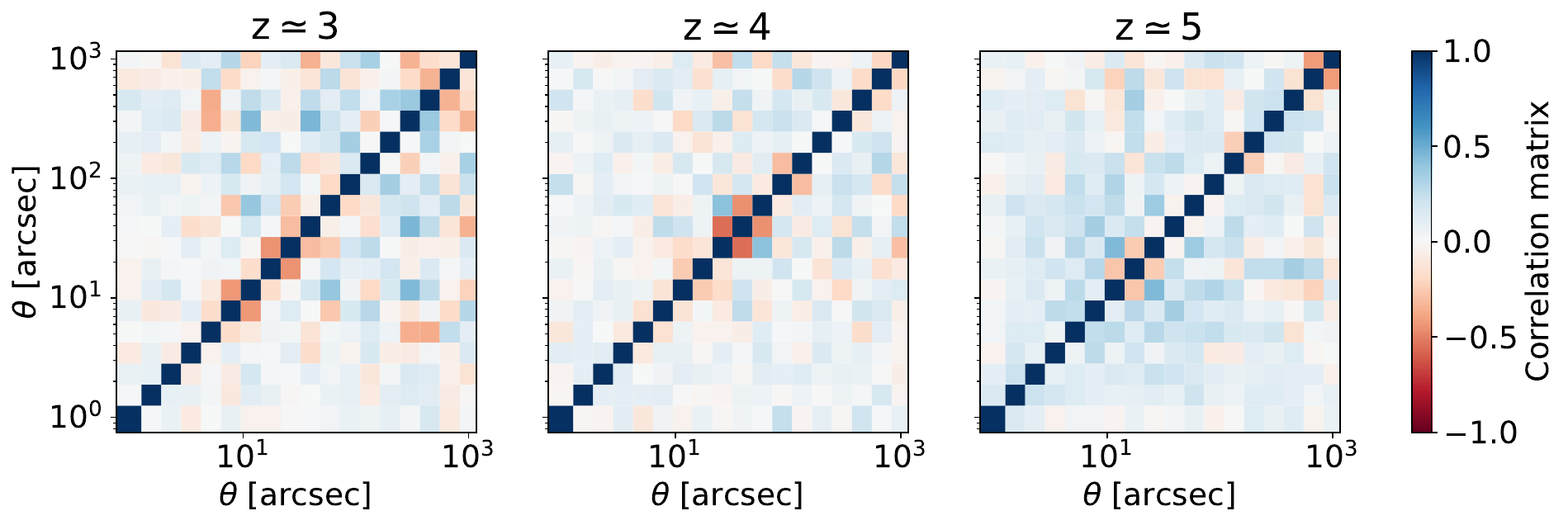}
    \caption{Measured correlation matrices for our $z\simeq3$, $z\simeq4$, and $z\simeq5$ full sample respectively, illustrating the correlation coefficients between different angular bins. The color scale represents the strength of correlation, with values ranging from -1 to 1. Each matrix is structured with axes representing angular separations $\theta$ in a logarithmic scale from 1 to $10^3$~arcsec. The correlation coefficients are computed from the covariance matrix using $\rho_{ij} = C_{ij} / \sqrt{C_{ii} \cdot C_{jj}}$ where $C$ is the covariance matrix, and $i,j$ are the angular bin indices. }
    \label{fig:all_z_correla_matrices}
\end{figure*}

Previous work has evaluated the performance of different error estimation techniques. \cite{Norberg_2009} compared the two error estimators, jackknife and bootstrap, on 1 $-$ 25 $h^{-1}$ Mpc scales, finding that the bootstrap method overestimated the variance of the two-point correlation function by $\sim40\%$ across all scales. In contrast, the jackknife method provided an accurate variance on large scales, but overestimated it on small scales. Overall, the jackknife estimator was found to be biased to a lesser extent.

In our analysis, we observe that the bootstrap error is larger than the jackknife error on smaller scales ($\theta < 4$~arcsec) but becomes smaller than the jackknife error for $\theta > 4$~arcsec. At scales of $\theta > 100$~arcsec, the bootstrap error is significantly smaller, potentially underestimating the error at the largest scales. As \cite{10.1093/mnras/stad3088} points out, bootstrap resampling randomly removes sources and therefore cannot account for systematic trends or variations in source density across the field, which may lead to an underestimation of errors in the correlation function. Considering these factors, we use only the jackknife error results in this paper.

We use \textsc{TreeCorr} \citep{Jarvis_2015} to calculate the correlation functions, but compute our own covariance matrices to have better control, while using the jackknife sky patches defined by \textsc{TreeCorr}.

\subsection{ACF measurements at $\mathbf{z\simeq3-5}$}
\label{subsec:2pt_func_result}

In Fig.~\ref{fig:w_all} (left), we present the angular correlation functions $w(\theta)$ measured for all galaxies in our three tomographic redshift bins within the COSMOS fields. Fig.~\ref{fig:all_z_correla_matrices} shows the correlation matrices for the full sample ACF in the three different redshift bins, for the same separation bins as in Fig.~\ref{fig:w_all}. At all redshifts, the correlation function shows an approximately power-law decline as $\theta$ increases, indicating a weakening of clustering at larger separations, as expected. As we move to higher redshifts, $z\simeq4$ and $z\simeq5$, the overall amplitude of $w(\theta)$ increases. This is largely due to the fact that we are observing only the brightest galaxies at these redshifts. These luminous galaxies tend to reside in more massive haloes \citep{2004MNRAS.353..189V}, which are linked to a higher galaxy bias \citep{Tinker_2010}. Hence, they are more clustered. The covariance matrices show mostly only mild correlations between nearby angular separation bins, with the $z \simeq 5$ sample showing a more significant pattern of correlation on separations of tens of arcsec.

We show the full ACFs (i.e., with no $M_{\rm{UV}}$ cut) as a demonstration of the maximum SNR we can obtain. The relative uncertainties tend to increase with $\theta$ and become larger at higher redshifts as expected. For angular separations ranging from $\theta \approx 1-300$ arcsec, the signal-to-noise ratios (SNRs) are high enough to yield useful information on clustering. However, at larger separations of $\theta \approx 300-1000$ arcsec, the SNRs are low, resulting in reduced statistical significance of the observed correlations. Also notable is the apparent steepening of the correlation function at the smallest separations; we discuss this further in Section~\ref{subsec:mcmc_fitting_testing}.

Since the luminosity cuts in different redshift bins are not the same, it is hard to interpret any evolution from the measured amplitude of the galaxy clustering from these ACFs. Hence, we applied a uniform $M_{\rm{UV}}$ cut to the sample and then re-measured the angular correlation functions to facilitate comparison across redshifts. The right panel of Fig.~\ref{fig:w_all} shows the resulting correlation functions of the three galaxy samples with an absolute magnitude cut of $M_{\rm{UV}}\le-20.0$. After applying the cut, a large portion of the faint-end galaxies at $z\simeq3$ were removed. For $z\simeq4$ and $z\simeq5$, the distributions are similar and mostly unaffected, with only a small fraction of the faint-end tail below the completeness limit being excluded. The correlation functions of the resulting samples are generally consistent, but show a slight evolution with redshift, most clearly seen around $\theta\lesssim60$~arcsec.

\section{Halo model fitting}
\label{sec:halo_model_fitting}

This section describes the halo model fitting process using \textsc{CCL}\footnote{\url{https://github.com/LSSTDESC/CCL}}\citep{Chisari_2019}, including the implementation of the 1-halo and 2-halo terms using HOD parameters for constructing the model angular two-point correlation function. We use a Bayesian MCMC approach to determine the posterior distribution of the HOD parameters. Various parameter combinations are tested, and we discuss the influence of specific parameters on fit quality. The MCMC posteriors are presented for different redshifts for both the full and magnitude-limited galaxy samples. Finally, we discuss the result of the inferred galaxies bias and duty cycle.

\subsection{The halo model}
\label{subsec:halo_model}

In the halo model, the power spectrum is modeled as the sum of terms that separately treat correlations between galaxies within each dark matter halo (the 1-halo term), and between pairs of haloes (the 2-halo term),
\begin{equation}
    P_{\text{full}}(k) = P_{1\text{-halo}}(k) + P_{2\text{-halo}}(k). 
	\label{eq:P_full}
\end{equation}
For the galaxy-galaxy halo power spectrum \citep{Nicola_2020}:
\begin{equation}
P_{1\text{-halo}}(k) = \frac{1}{\bar{n}_g^2} \int \frac{dn}{dM} dM \, \bar{N}_c \, f_c \left[ \bar{N}_s^2 u_s^2(k) + 2 \bar{N}_s u_s(k) \right],
\end{equation}
\begin{equation}
P_{2\text{-halo}}(k) = \left( \frac{1}{\bar{n}_g} \int \frac{dn}{dM} dM \, b_h(M) \bar{N}_c \left[ f_c + \bar{N}_s u_s(k) \right] \right)^2 P_{\text{lin}}(k).
\end{equation}
Here, $M$ is the halo mass, $\frac{dn}{dM} $ is the halo mass function, which is the number density of dark matter haloes per unit mass interval, and $u_s(k)$ is the Fourier transform of the normalised density profile of satellite galaxies. $f_c$ is a parameter explained later in Equation~\ref{eq:ng_r_M}. We use the \cite{Tinker_2008} model as our halo mass function and the \cite{Tinker_2010} result for the halo bias $b_h(M)$. $\bar{n}_g$ is the total mean galaxy density. 
The galaxy bias $b^2_g(k, z)$ is defined as
\begin{equation}
b^2_g(k, z) = \frac{P_{\text{full}}(k)}{P_{\text{lin}}(k)},
\label{eq:galaxy_bias}
\end{equation}
where $P_{\text{lin}}(k)$ is the linear matter power spectrum. In the large-scale limit where $k\rightarrow0$, the full power spectrum $P_{\text{full}}$ is represented by the 2-halo term power spectrum $P_{2\text{-halo}}(k)$ only.

The HOD \citep{cooray_2002} is a key component of the halo model power spectrum calculation.
In \textsc{CCL}, the mean observed galaxy density profile within a halo of mass M is modelled as
\begin{equation}
\langle n_g(r)|M \rangle = \bar{N}_c(M) \left[f_c + \bar{N}_s(M) u_{s}(r|M)\right],
\label{eq:ng_r_M}
\end{equation}
where $f_c$ is the observed fraction of central galaxies (between 0 and 1). $\bar{N}_c(M)$ is the mean number of central galaxies as a function of halo mass $M$, defined as

\begin{equation}
\bar{N}_c(M) = \frac{1}{2} \left[ 1 + \text{erf} \left( \frac{\log(M/M_{\rm min})}{\sigma_{\ln M}} \right) \right],
\label{eq:N_c}
\end{equation}
where $\text{erf}$ is the error function, $M_{\text{min}}$ is the minimum mass threshold for hosting one central galaxy, and $\sigma_{\ln M}$ is the scatter in the logarithm of the halo mass.

Similarly, the mean number of satellite galaxies (with one central galaxy), $\bar{N}_s(M)$, is given by
\begin{equation}
\bar{N}_s(M) = \Theta(M - M_0) \left( \frac{M - M_0}{M_1} \right)^\alpha,
\label{eq:N_s}
\end{equation}
where $\Theta(x)$ is the Heaviside step function, which ensures that only haloes with mass greater than $M_0$ have satellite galaxies. Here, $M_1$ is a normalisation mass scale, and $\alpha$ is the slope of the relation.

The normalised distribution of satellites, denoted as $u_{s}(r|M)$, depends on the radial distance $r$ from the center of the halo. In this work we assume it follows the Navarro, Frenk, $\&$ White (NFW) profile \citep{Navarro_1997}
\begin{equation}
u_s(r|M) \propto \Theta(r_{\rm max} - r) \left( \frac{r/r_g}{(1 + r/r_g)^2} \right).
\label{eq:us_r_M}
\end{equation}
Here, $\Theta(r_{\rm max} - r)$ ensures that the profile is defined only within a maximum radius $r_{\rm max}$. The free parameters in our model are $\log_{10} M_{\rm min}$, $\log_{10} M_1$, $\alpha$, $f_c$ and $\beta_g$.

The volume integral of the satellite distribution $u_s$ is 1. The characteristic radius $r_g$ is linked to the NFW scale radius $r_s$ by the relation $r_g = \beta_g r_s$, where $\beta_g$ is a proportionality constant, which is one of the free parameters in our halo model fitting. The radius $r_{\rm max}$ is connected to the overdensity radius $r_\Delta$ through the expression $r_{\rm max} = \beta_{\rm max} r_\Delta$. The scale radius itself is related to the comoving overdensity halo radius by the concentration-mass relation, expressed as $r_\Delta(M) = c(M) r_s$, where $c(M)$ is the concentration parameter as a function of mass. In our analysis, we use the concentration-mass relation result from \cite{Bhattacharya_2013}.

One can use the halo model power spectrum and a tracer $\Delta_\ell^i(k)$ to construct the angular power spectrum $C_\ell$:
\begin{equation}
    C_\ell^{ij} = 4\pi \int_0^\infty \frac{dk}{k} P_\Phi(k) \Delta_\ell^{i}(k) \Delta_\ell^{j}(k),   
	\label{eq:}
\end{equation}
where, in our case, $\Delta_\ell^{i}(k) = \Delta_\ell^{j}(k) = \Delta^{NC}$, where $\Delta^{NC}$ is the galaxy number counts tracer. Galaxies form in the haloes where there is an overdensity of dark matter. Therefore, the number count of galaxies serves as a biased tracer of the underlying matter density field. The number count transfer function $\Delta_\ell^{NC}(k)$ describes how primordial fluctuations in $k$, $\mathcal{P}_\Phi(k)$, are transferred into the observed fluctuation in number counts at multiple $l$:
\begin{equation}
   \Delta_\ell^{NC}(k) = \int dz \, p_z(z) b_g(k, z) T_\delta(k, z) j_\ell(k \chi(z)).  
	\label{eq:Delta_ell}
\end{equation}
Here, $T_\delta$ is the matter density transfer function. $b_g(k, z)$ is the galaxy bias (Equation~\ref{eq:galaxy_bias}). $p_z(z)$ is the normalised redshift distribution of sources. $j_\ell(\chi)$ is the $\ell-$th order spherical Bessel function, which projects the 3D fluctuations onto the celestial sphere. $\chi(z)$ is the comoving distance to redshift $z$.

To obtain the angular correlation function $w$ from $C_{\ell}$, the equation shows the $w^{ij}$ for two bins $i$ and $j$ as a sum over all multiple moments $\ell$:
\begin{equation}
    w^{ij}(\theta) = \sum_{\ell} \frac{2\ell + 1}{4\pi} (-1)^{\ell} C_{\ell}^{NN} d_{s_N, \pm s_N}^{\ell}(\theta).  
	\label{eq:w_ij}
\end{equation}
$w^{ij} =w^{NN}$ denotes the correlation function related to the angular power spectra for two galaxy number density fields that are denoted by the $N$s. $d_{s_N, s_N}^{\ell}(\theta)$ is the Wigner-d matrices \citep{Ng_1999,Chon_2004}, which describe the projection of the 3D fluctuations onto the celestial sphere. $s_N=0$ is the spin of the galaxy number density field.

\subsection{MCMC fitting method}
\label{subsec:mcmc_method}

In this study, we fit an HOD model to the measured angular two-point correlation function using the MCMC method. We use the \textsc{emcee} package from \cite{Foreman_Mackey_2013}. The likelihood function for this fit is expressed as:
\begin{equation}
\begin{aligned}
\log \mathcal{L} = -\frac{1}{2} \biggl\{ & \sum_{i,j} \left[ w_{\text{obs}}(\theta_i) - w(\theta_i; \vec{p}) \right] C_{ij}^{-1} \left[ w_{\text{obs}}(\theta_j) - w(\theta_j; \vec{p}) \right] \\
& + \frac{\left[ \log n_g^{\text{theor}}(\vec{p}) - \log n_g^{\text{obs}} \right]^2}{\sigma_{\log n_g(\vec{p})}^2} \biggr\},
\label{eq:log_likelihood}
\end{aligned}
\end{equation}
where $w_{\text{obs}}(\theta_i)$ represents the observed angular correlation function at angular separation $\theta_i$, with $i$, $j$ denoting angular bins. $C_{ij}$ is the full covariance matrix of the observed ACF. The model correlation function $w(\theta_i; \vec{p})$ and the theoretical number density $n_g^{\text{theor}}(\vec{p})$ are functions of the HOD parameters, where $\vec{p}$ is a vector of HOD parameter values. $n_g^{\text{obs}}$ is the observed galaxy number density of our data, and $\sigma_{\log n_g}$ is the standard deviation in $\log n_g$, which we choose to be a conservative value of $10\%$ of $\log n_g^{\text{obs}}$, following \cite{Harikane_2022}. We use the logarithm of $n_g$ because otherwise the fitting tends to overly prioritise matching the number density, resulting in poor fits to the ACF and poorly constrained posterior distributions. Given the number of galaxies $N$ within the comoving volume $V_{C}$, the number density $n_g$ is: $n_g = N/V_{C}$. The model correlation function is defined as:
\begin{equation}
w(\theta_i; \vec{p}) = w_{\text{theor}}(\theta_i; \vec{p}) - \text{IC} + \frac{1}{N},
\label{eq:w_theor}
\end{equation}
where $w_{\text{theor}}$ is the theoretical prediction of the correlation function from the halo model, as described in Section \ref{subsec:halo_model}. As the effects of shot noise and the finite survey volume through the integral constraint (IC) are present in the observed correlation function but not in the theoretical correlation function, we accounted for them in the fitting. The shot noise term is given by $1/N$, where $N$ is the total number of galaxies per radian-squared within the redshift bin. Since the correlation function must satisfy the condition that its integral over the survey volume vanishes, the IC applies a constant correction to the modelled correlation function (e.g. \citealt{Groth_1977, Roche_1999, Riquelme_2023}). To meet the condition, we calculate the IC as follows:
\begin{equation}
\text{IC} = \frac{\sum_{i}N_{rr}\,w_{\text{obs}}(\theta_i)}{N_{rr}}.
\label{eq:IC}
\end{equation}

We have five free parameters from the HOD model: $\log_{10}M_{\text{min}}$, the logarithm of the minimum mass of a halo that can host a central galaxy; $\log_{10}M_1$, the logarithm of the characteristic halo mass at which on average one satellite galaxy is found per halo; $\alpha$, the slope of the power law describing the number of satellite galaxies as a function of halo mass; $f_c$, the observed fraction of central galaxies; and $\beta_g$, which is related to the spatial distribution of satellite galaxies within their host haloes. We set $\sigma_{\ln{M}} = 0.2$, which controls the scatter between the halo mass and the minimum mass for galaxy occupation. This is a characteristic transition width, and its specific value is not critical as long as it remains relatively small. Additionally, we set $\log_{10}M_{0} = \log_{10}M_{\text{min}}+0.5$, for setting a baseline for the satellite galaxies, assuming that $\log_{10}M_{0}$ should be slightly larger than $\log_{10}M_{\text{min}}$. Both values are based on \cite{Harikane_2022} with slight modification. The parameters are assumed constant in redshift within each of our redshift bins, but are fitted separately for each bin.

The priors for the parameters in this study are all specified as uniform distributions, defined as follows. The parameter $\log_{10}M_{\text{min}}$ is constrained to be within the range of $0.0$ to $15.0$, while $\log_{10}M_1$ has a slightly larger range, from $0.0$ to $17.0$, with the additional condition that $\log_{10}M_1$ must be greater than $\log_{10}M_{\text{min}}$. The parameter $\alpha$ is constrained to values greater than or equal to $0.0$, $f_c$ is bounded between $0.0$ and $1.0$, and $\beta_g$ is bounded between $0.0$ and $4.0$. We choose very wide priors, except for restricting $f_c$ to be a fraction between 0 and 1. We use 64 walkers for the MCMC process, and run the chains for 3000 steps. We tested discarding the first 10$\%$ and 20$\%$ of the chain as burn-in and found that the mean values changes by $\leq$ $0.05\sigma$, and inspected some of the chains visually to ensure reasonable convergence. We kept the final 80$\%$ of each chain for our analysis.

\begin{figure}
    \centering
    \includegraphics[width=\linewidth]{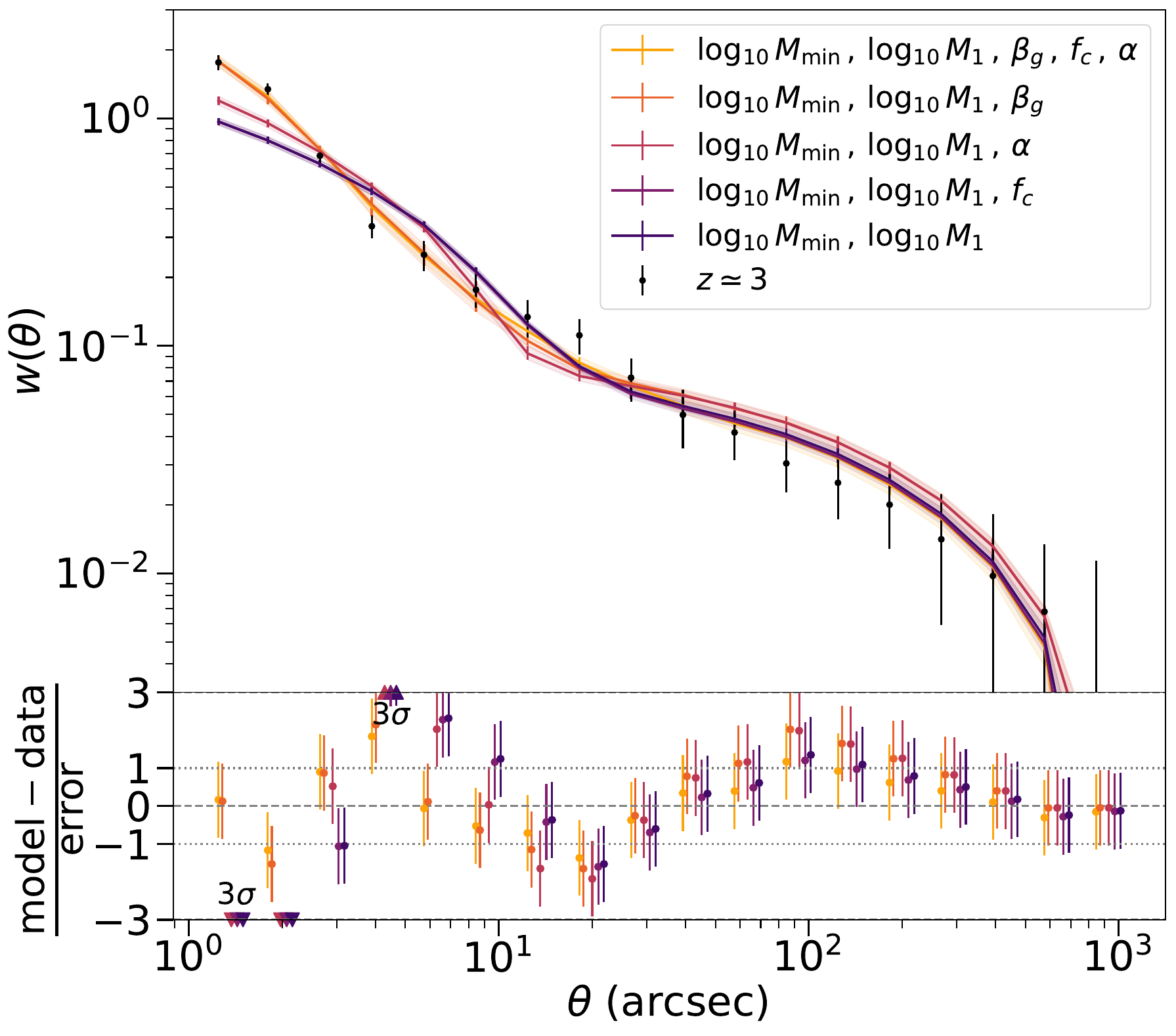}
    \caption{Halo model fitting results for the angular correlation function $w(\theta)$ at $z\simeq3$, comparing the observational data (dots with error bars) to several theoretical models (solid lines) with different combinations of free parameters. The different colours correspond to different sets of free parameters used in the fits. If the parameter is not free, it is fixed to 1. The solid lines represent the model fits, and the shaded regions and error bars correspond to the $1\sigma$ uncertainties in the model predictions. We apply small horizontal shifts in the lower panel for clarity. The lower panel shows the z-score, defined as the difference between the model and the data, normalised by the error bars. The horizontal dashed lines indicate the $1\sigma$ and  $3\sigma$ significance levels, with the triangular markers indicating data points that exceed a $3\sigma$ deviation.}
    \label{fig:halo_model_z3}
\end{figure}

\subsection{Testing different combinations of model parameters}
\label{subsec:mcmc_fitting_testing}

To test the behaviour of the halo model parameter inference procedure, we experimented with different combinations of the free parameters $\log_{10}M_{\text{min}}$, $\log_{10}M_1$, $\beta_g$, $f_c$, $\alpha$ from the HOD model in \textsc{CCL} over the redshift range from $z=2.75$ to $z=3.5$, alternately fixing some and allowing others to vary.

On very small scales within dark matter haloes, \cite{Masjedi_2006} reported steep clustering that deviates from predictions of the standard HOD model, and that is likely driven by physical processes. While they mentioned that modifying HOD models by allowing a galaxy distribution more concentrated than dark matter could potentially explain such enhanced clustering, our HOD model has the free parameter $\beta_g$, a dimensionless scaling constant relating the galaxy distribution radii to the dark matter halo radii (see Section \ref{subsec:halo_model}). While adjusting the $\beta_g$ parameter could effectively reproduce the steep clustering at very small scales and provide good fits, we intentionally exclude the smallest-scale data points to minimise potential bias from this effect.

Recent observations by \cite{Duan_2024} presented observational evidence of enhanced star formation in galaxy pairs with separations smaller than approximately 20~kpc at high redshifts ($z=4.5-8.5$). 
This scale of $\sim20$ kpc is chosen as a cutoff in this study, as this is approximately the scale at which the galaxy-galaxy interactions could begin to affect the observed number of LBGs, particularly by boosting star formation rates and hence observed UV magnitude, which is not captured by the simple HOD model (see Appendix~\ref{appendixpairs} for further discussion). 
We calculate the angular separations corresponding to a physical distance of 20~kpc at our redshifts of interest, and subsequently exclude data points measuring separations smaller than these angular scales from our fitting analysis. We apply a minimum angular separation cut of 2.57, 2.81, 3.07 arcsec at $z\simeq3, 4, 5$ respectively. This criterion results in omitting exactly the first three data points at each redshift.

To illustrate the impact of this cut, we perform a test in which all data points are included in the fit for the $z\simeq3$ full sample. Five parameters are available to be varied: $\log_{10} M_{\rm min}$, $\log_{10} M_1$, $\alpha$, $f_c$, $\beta_g$. The parameters $\log_{10}M_{\rm min}$ and $\log_{10}M_1$ are always varied; any parameter not varied is fixed to 1. The purpose of these tests is to identify the combination of free parameters that provides the best fit. Recall that the fit is constrained by both the observed angular correlation function and the observed galaxy number density (see Equation~\ref{eq:log_likelihood}). Fig.~\ref{fig:halo_model_z3} shows the results of the MCMC halo model fitting applied to the ACFs for the full sample under these conditions. We present the one-dimensional posterior distributions of the model parameters for the full and magnitude-limited samples at redshift $z\simeq3$ in Table \ref{tab:MCMC_parameter_vari}, with variations in the choice of free parameters.

\begin{figure}
    \centering
    \includegraphics[width=\linewidth]{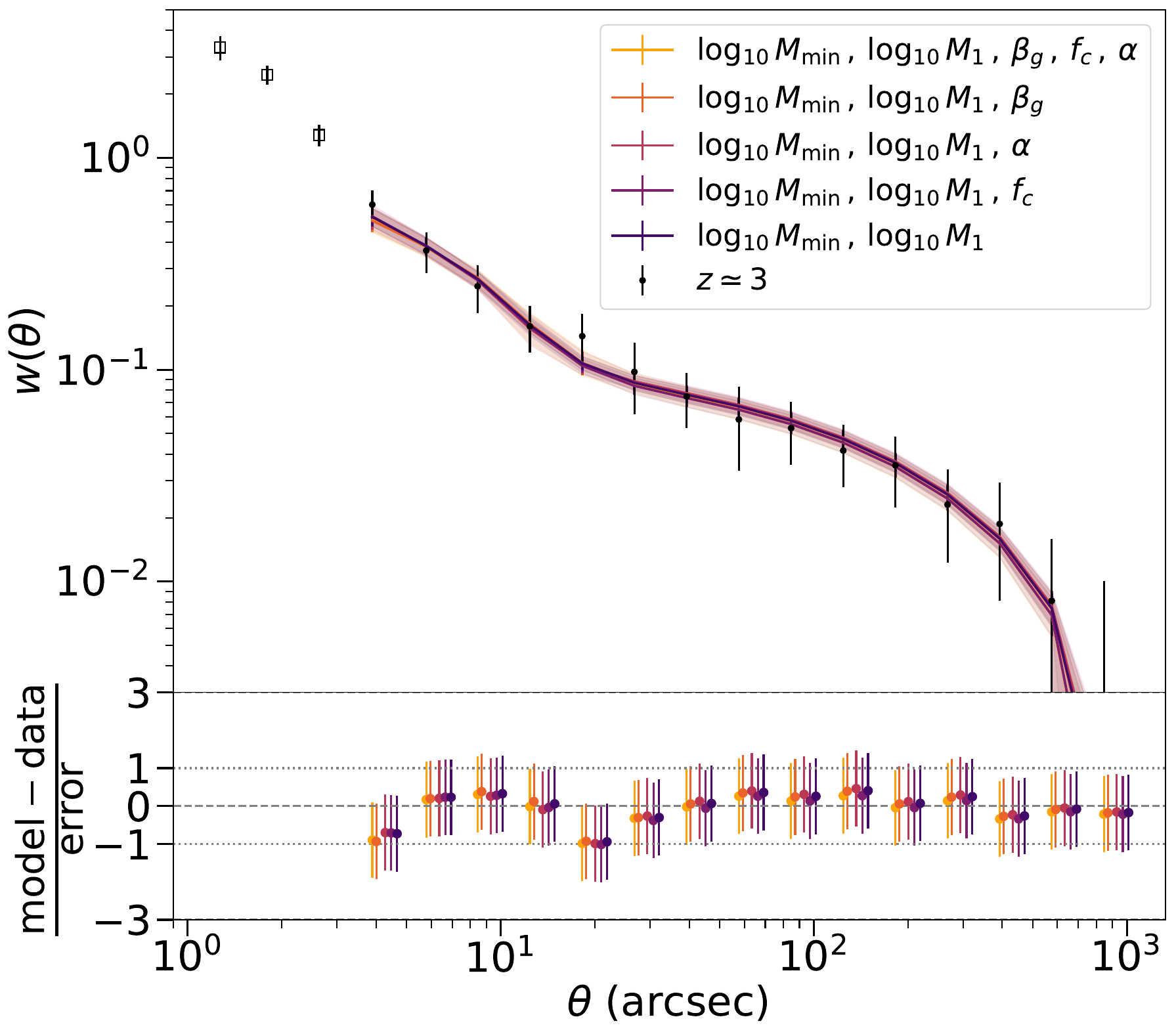}
    \caption{The same as Fig.~\ref{fig:halo_model_z3}, but excluding the first three data points (displayed as open squares) at the smallest angular scales for the $M_{\rm{UV}}$-limited sample at $z\simeq3$.}
    \label{fig:halo_model_z3_magcut}
\end{figure}

\begin{table*}
    \renewcommand{\arraystretch}{1.4}
    \centering
    \begin{tabular}{c c c c c c c c c c c}
        \hline
        \hline
        $z$ & $\log_{10} (M_{\rm min} / M_\odot)$ & $\log_{10} (M_{1} / M_\odot)$ & $\alpha$ & $f_c$ & $\beta_g$ & $n_g^{\text{theor}} \, (10^{-3}{\rm Mpc}^{-3})$ & $n_g^{\text{obs}} \, (10^{-3}{\rm Mpc}^{-3})$ & $\chi^2_{\nu}$ & $\nu$ & \text{Bias} \\
        \hline
        \multicolumn{11}{l}{Full sample} \\
        \hline
        3 & $11.22^{+0.07}_{-0.07}$ & $12.22^{+0.10}_{-0.11}$ &  -- & -- & $0.07^{+0.04}_{-0.04}$ & $4.23^{+1.07}_{-0.83}$ & 3.11 & 4.21 & 16 & $2.85^{+0.08}_{-0.07}$ \\
        3 & $11.02^{+0.09}_{-0.08}$ & $12.12^{+0.23}_{-0.17}$  & -- & $0.54^{+0.25}_{-0.25}$ & -- & $4.28^{+2.07}_{-1.91}$ & 3.11 & 10.35 & 16 & $2.68^{+0.07}_{-0.06}$ \\\
        3 & $11.26^{+0.05}_{-0.08}$ & $12.63^{+1.29}_{-0.52}$ & $0.33^{+0.21}_{-0.17}$ & -- & -- & $3.84^{+1.19}_{-0.65}$ & 3.11 & 9.19  & 16 & $2.84^{+0.07}_{-0.08}$ \\
        3 & $11.06^{+0.08}_{-0.07}$ & $11.92^{+0.11}_{-0.11}$ & -- & -- & -- & $7.26^{+1.88}_{-1.49}$ & 3.11 & 9.80 & 17 & $2.71^{+0.06}_{-0.06}$ \\
        3 & $11.13^{+0.07}_{-0.08}$ & $12.82^{+0.30}_{-0.21}$ & $2.35^{+0.89}_{-0.47}$ & $0.70^{+0.21}_{-0.25}$ & $0.01^{+0.01}_{-0.00}$ & $3.14^{+1.38}_{-1.15}$ & 3.11 & 1.70 & 14 & $2.67^{+0.10}_{-0.10}$ \\
        \hline
        \multicolumn{11}{l}{$M_{\rm UV}$-limited sample} \\
        \hline
        3 & $11.50^{+0.09}_{-0.10}$ & $12.54^{+0.15}_{-0.16}$ &  -- & -- & $1.84^{+1.37}_{-1.17}$ & $1.74^{+0.65}_{-0.44}$ & 1.13 & 0.34 & 13 & $3.23^{+0.13}_{-0.13}$ \\
        3 & $11.46^{+0.10}_{-0.11}$ & $12.74^{+0.24}_{-0.22}$  & -- & $0.59^{+0.27}_{-0.27}$ & -- & $1.11^{+0.63}_{-0.45}$ & 1.13 & 0.31 & 13 & $3.18^{+0.14}_{-0.14}$ \\\
        3 & $11.53^{+0.08}_{-0.11}$ & $12.68^{+0.85}_{-0.22}$ & $0.86^{+0.45}_{-0.52}$ & -- & -- & $1.58^{+0.65}_{-0.37}$ & 1.13 & 0.31  & 13 & $3.26^{+0.12}_{-0.13}$ \\
        3 & $11.50^{+0.09}_{-0.09}$ & $12.54^{+0.15}_{-0.16}$ & -- & -- & -- & $1.74^{+0.64}_{-0.44}$ & 1.13 & 0.32 & 14 & $3.23^{+0.13}_{-0.12}$ \\
        3 & $11.48^{+0.10}_{-0.12}$ & $12.83^{+0.75}_{-0.26}$ & $0.98^{+0.43}_{-0.55}$ & $0.62^{+0.25}_{-0.26}$ & $1.76^{+1.50}_{-1.29}$ & $1.12^{+0.59}_{-0.44}$ & 1.13 & 0.14 & 11 & $3.18^{+0.14}_{-0.14}$ \\
        \hline
        \hline
    \end{tabular}
    \caption{Halo model MCMC fitting posteriors for varying free parameters at $z\simeq3$. It presents median values and $1\sigma$ uncertainties obtained from MCMC fitting of a halo model to galaxy data at $z\simeq3$. The upper section of the table  shows results from fitting to the full galaxy sample (Fig.~\ref{fig:halo_model_z3}), and the lower section shows the results obtained from fitting to the he $M_{\rm{UV}}$-limited galaxy sample (Fig.~\ref{fig:halo_model_z3_magcut}). For each fit, the predicted number density $n_{g}^{\text{theor}}$ and the galaxy bias (bias) derived from the model are provided, along with the observed galaxy number density $n_{g}^{\text{obs}}$ for reference. The reduced chi-square $\chi^2_{\nu} \equiv \chi^2 / {\rm d.o.f.}$, where the number of degrees of freedom $\nu$ is calculated as the number of correlation function data points plus one (for the measured $n_g$), minus the number of free parameters in the model. Parameters shown as dashes in the table are fixed to a value of 1.}
    \label{tab:MCMC_parameter_vari}
\end{table*} 

Testing on the full sample is motivated by the desire to understand how well the model can reproduce the ACF without omitting any data points. Among the tested models, the best-fitting one includes all five free parameters. This model has the lowest reduced $\chi^2$ value $\chi^2_\nu$, reflecting the best overall agreement with both the measured ACF and the observed number density $n_{g}^{\text{obs}}$. The second best-fitting model, which includes the parameters $\log_{10}M_{\rm min}$, $\log_{10}M_1$, $\beta_g$, also performs relatively well, though it yields a higher $\chi^2_\nu$ value. The higher $\chi^2_\nu$ arises from a worse match to both the ACF and the number density. The remaining models, which do not include $\beta_g$ as a free parameter (i.e. with $\beta_g=1$), perform much worse in the fit.

If $\beta_g$ is fixed to 1 (as in the remaining cases), the satellite galaxy distribution has a profile with a scale radius at the dark matter halo's scale radius $r_s$ (Equation~\ref{eq:us_r_M}). Note that when $\beta_g$ is set as free and fitted for all observed ACF data points, the fitted results are very small ($\beta_g \sim 0.01$) for the three redshift bins. This indicates that the satellite galaxies are modeled as being extremely close to the central of the halo.

\begin{figure*}
    \centering
    \includegraphics[width=0.98\linewidth]{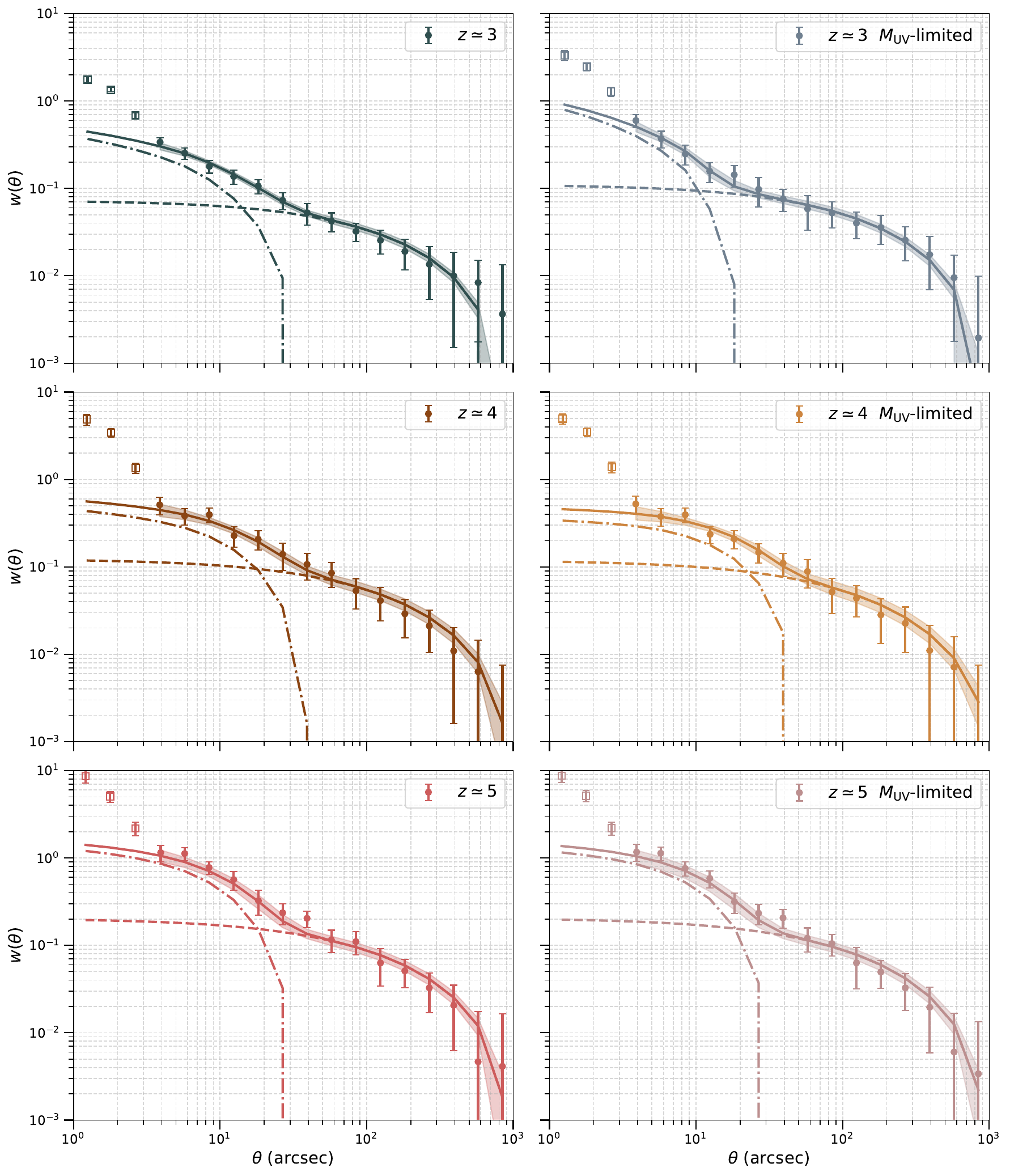}
    \caption{Angular correlation functions for luminous LBGs at redshifts $z\simeq3$, $z\simeq4$, and $z\simeq5$. The left column shows the measurements for the full galaxy sample, the right column corresponds to magnitude limited subsample with $M_{\rm{UV}}\le-20.0$. The data points with error bars represent the measured correlation functions, with the first three data points at the smallest angular scales displayed as open squares and excluded from the fitting. The solid, dashed, and dash-dotted lines correspond to the theoretical predictions from the full halo model, the 2-halo term, and the 1-halo term, respectively. The halo model terms are extended to the smallest angular scale for reference based on the fit to the solid data points. The shaded regions around the solid lines indicate the uncertainty in the halo model predictions: the solid lines represent the 50th percentile of the MCMC posterior, and the shaded regions mark the 16th and 84th percentiles. The full halo model describes the clustering of galaxies by accounting for both the 1-halo term (galaxies within the same halo) and the 2-halo term (galaxies in separate haloes). In our analysis, the halo model for fitting includes the shot noise present in the measured angular correlation function and the subtraction of the IC, see Section \ref{subsec:mcmc_method} for details. All axes are displayed on a logarithmic scale.}
    \label{fig:angular_correlation}
\end{figure*}

The third-best case is the one where $\alpha$ is the only additional free parameter to $\log_{10}M_{\rm min}$ and $\log_{10}M_1$. It controls how rapidly the number of satellite galaxies increases with halo mass. Allowing $\alpha$ to vary mainly affects smaller angular scales, where the contribution from satellite galaxies is significant. However, with $\beta_g=1$, we can tell from the figure that varying the number of satellite galaxies alone cannot fully compensate for the fit.

The inclusion of $f_c$ while keeping $\alpha=1$ makes only a marginal difference in the fits. It represents the observed fraction of central galaxies and is in fact a very close concept to the duty cycle. The duty cycle is the fraction of time that the galaxies are observable in our data. Some LBGs may only be detectable during temporarily luminous phases (e.g. a starburst), as they are detected based on a rest-frame UV light. We can treat this factor as the duty cycle by slightly modifying Equation~\ref{eq:ng_r_M}, taking the $f_c$ as a multiplicative factor. The value of $f_c$ remains the same, but leads to a redefinition of $M_1$ as
\begin{equation}
    M_1 \to M_1 f_c^\frac{1}{\alpha}.
    \label{eq:M1_NEW}
\end{equation}
In the main results of this work, we omit the first three data points at the smallest angular scales, as discussed earlier in this section. To ensure a consistent analysis, we repeat the same parameter tests for the magnitude-limited subsample ($M_{\rm{UV}}\le-20.0$) with the first three ACF data points excluded from the fitting (Fig.~\ref{fig:halo_model_z3_magcut}). The lower portion of Table~\ref{tab:MCMC_parameter_vari} shows the corresponding one-dimensional posterior distributions of the model parameters. 

After excluding the first three data points, the inclusion or exclusion of individual parameters no longer significantly affects the fit quality (as seen in the lower panels of Fig.~\ref{fig:halo_model_z3} and Fig.~\ref{fig:halo_model_z3_magcut}). The model with all five parameters free still provides the best agreement with the data, but the differences in fit quality between models are very small, as all predicted curves closely follow the measured clustering signal. This also indicates that fixing or freeing parameters can yield the same goodness of fit, while resulting in different model parameter values for $z\sim 3$. For cases where $\beta_{g}$ is freed, the fitted values no longer hits the $\beta_g>0$ prior, further suggesting that the signal is well described by the halo model.

It is worth noting that, although the fitted values of $\alpha$ and $\beta_g$ are consistent with the fiducial values of 1, the inclusion of these parameters impacts the fitting results. For example, when $\alpha$ is fixed to 1, the constraints on $\log_{10}M_1$ are much smaller than the case where $\alpha$ varies. Furthermore, as we show later in Section \ref{subsec:subsec:mcmc_fitting_result}, for other redshift bins, the fitted values of $\alpha$ can deviate from 1 with high statistical significance. Additionally, the $\chi^2_\nu$ values are all very small, which suggests an over fitting issue. Nevertheless, since there is no physical justification for fixing any of the parameters and we are able to constrain all five parameters with the available data (see Section~\ref{subsec:subsec:mcmc_fitting_result}), we adopt the five-parameter model for the final fit.

\subsection{Results of halo model fits}
\label{subsec:subsec:mcmc_fitting_result}
 
The solid lines in Fig.~\ref{fig:angular_correlation} show the halo model predictions for the galaxy ACF across the three different redshifts when excluding the first three observed data points for fitting. The solid line reflects the combined contributions of the 1-halo and 2-halo terms. The halo model predictions are extended beyond the fitted solid points at the smallest scales for reference, showing the level of clustering amplitude expected. Table~\ref{tab:MCMC_parameter_results} presents the one-dimensional posterior distributions of the model parameters for the three different redshift bins and two sample selections (full and $M_{\rm{UV}}$-limited samples). Corner plots for these parameters are shown in Figs.~\ref{fig:contourplot_z3}, \ref{fig:contourplot_z4}, and \ref{fig:contourplot_z5} in Appendix~\ref{corner_plots}, illustrating the two-dimensional posterior distributions for the $z\simeq3$, $z\simeq4$, and $z\simeq5$ full sample fits, respectively. 

\begin{table*}
    \renewcommand{\arraystretch}{1.4}
    \centering
    \begin{tabular}{c c c c c c c c c c c}
        \hline
        \hline
        $z$ & $\log_{10} (M_{\rm min} / M_\odot)$ & $\log_{10} (M_{1} / M_\odot)$ & $\alpha$ & $f_c$ & $\beta_g$ & $n_g^{\text{theor}} \, (10^{-3}{\rm Mpc}^{-3})$ & $n_g^{\text{obs}} \, (10^{-3}{\rm Mpc}^{-3})$ & $\chi^2_{\nu}$ & $\nu$ & \text{bias} \\
        \hline
        $3^{\ast}$ & $10.97^{+0.10}_{-0.12}$ & $12.43^{+0.20}_{-0.17}$ & $1.58^{+0.27}_{-0.16}$ & $0.46^{+0.29}_{-0.21}$ & $1.79^{+1.50}_{-1.39}$ & $3.89^{+2.25}_{-1.58}$ & 3.11 & 0.23 & 11 & $2.60^{+0.09}_{-0.10}$ \\
        $4^{\ast}$ & $11.25^{+0.12}_{-0.15}$ & $12.70^{+0.25}_{-0.18}$ & $2.42^{+0.91}_{-0.42}$ & $0.48^{+0.32}_{-0.24}$ & $1.74^{+1.50}_{-1.38}$ & $0.96^{+0.63}_{-0.44}$ & 0.759 & 0.23 & 11 & $3.70^{+0.23}_{-0.25}$ \\
        $5^{\ast}$ & $10.89^{+0.11}_{-0.13}$ & $12.39^{+0.32}_{-0.18}$ & $2.61^{+1.12}_{-0.43}$ & $0.37^{+0.34}_{-0.19}$ & $0.77^{+0.82}_{-0.61}$ & $1.08^{+0.82}_{-0.50}$ & 0.704 & 0.47 & 11 & $4.24^{+0.24}_{-0.25}$ \\
        \hline
        3 & $11.48^{+0.10}_{-0.12}$ & $12.83^{+0.75}_{-0.26}$ & $0.98^{+0.43}_{-0.55}$ & $0.62^{+0.25}_{-0.26}$ & $1.76^{+1.50}_{-1.29}$ & $1.12^{+0.59}_{-0.44}$ & 1.13 & 0.14 & 11 & $3.18^{+0.14}_{-0.14}$ \\
        4 & $11.19^{+0.15}_{-0.17}$ & $12.89^{+0.36}_{-0.22}$ & $3.18^{+1.58}_{-0.73}$ & $0.40^{+0.34}_{-0.22}$ & $1.39^{+1.67}_{-1.15}$ & $0.97^{+0.82}_{-0.46}$ & 0.750 & 0.34 & 11 & $3.58^{+0.27}_{-0.29}$ \\
        5 & $10.91^{+0.12}_{-0.13}$ & $12.40^{+0.26}_{-0.17}$ & $2.65^{+0.92}_{-0.40}$ & $0.39^{+0.31}_{-0.20}$ & $0.87^{+0.77}_{-0.68}$ & $1.05^{+0.76}_{-0.47}$ & 0.698 & 0.49 & 11 & $4.27^{+0.25}_{-0.26}$ \\
        \hline
        \hline   
    \end{tabular}
    \caption{Halo model MCMC fitting posteriors for $z\simeq3$, $z\simeq4$ and $z\simeq5$ measurements with the first three data points at the smallest scale omitted, corresponding to the results shown in Fig.~\ref{fig:angular_correlation}. We present median values and $1\sigma$ uncertainties obtained from MCMC fitting of halo model to galaxy data. The upper section of the table (above the horizontal line) shows results from fitting to the full galaxy sample, the lower section shows the results obtained from fitting to the $M_{\rm{UV}}$-limited galaxy sample. The fits are illustrated in Fig.~\ref{fig:angular_correlation}. The number densities $n_{g}^{\text{theor}}$ and $n_{g}^{\text{obs}}$, reduced chi-square, $\chi^2_{\nu}$, are defined as in Table~\ref{tab:MCMC_parameter_vari}. \newline $^{\ast}$ The redshift bins use different magnitude cuts: $z\simeq3$ ($m>27.1$), $z\simeq4$ ($m>26.9$), and $z\simeq5$ ($m>26.5$).}
    \label{tab:MCMC_parameter_results}
\end{table*} 

The halo model provides good fits to the ACFs at $z\simeq3$ and $z\simeq4$ for both the full and $M_{\rm{UV}}$-limited samples. At $z\simeq5$, the fits are less robust, particularly for the $M_{\rm{UV}}$-limited sample, where the model under-predicts the clustering amplitude at the 1-halo regime. For both samples, the halo mass threshold scales $M_{\rm min}$ are on the order of $10^{11}\textup{M}_\odot$, and the satellite halo masses $M_1$ are on the order of $10^{12}\textup{M}_\odot$. The $M_{\rm min}$ values are well constrained, as they heavily control the galaxy number density $n_{g}^{\text{theor}}$. Although $M_{1}$ is degenerate with $\alpha$, it is nevertheless reasonably constrained.

The full galaxy samples do not share the same absolute magnitude cut, so interpretations of redshift evolution in the fitted parameters should be treated with caution. For the $M_{\rm{UV}}$-limited samples, the evolution of galaxy bias and $f_c$ is discussed in Section~\ref{subsubsec:subsec:bias} and Section~\ref{subsubsec:subsec:duty_cycle}. The fitted values of the $\beta_g$ values are within a physically reasonable range but are weakly constrained. This is because the satellite galaxies have only a minor influence on the fit due to the first three data points at the smallest scales being excluded. 

Using the halo model parameters, we can reconstruct the theoretical number densities $n_g^{\text{theor}}$, which we use for matching the galaxy abundance as discussed in Section~\ref{subsec:mcmc_method}. As shown in Table~\ref{tab:MCMC_parameter_results}, the posterior of $n_g^{\text{theor}}$ gives a large error bar which is intended by our abundance matching formalism. The wide posterior of $n_g^{\text{theor}}$ suggests that most of our constraining power comes from the angular correlation function, instead of a strict matching of galaxy number density.

Overall, the model successfully captures the clustering signals across all scales at all three redshift bins. The intermediate scale region is particularly well fitted at $z\simeq3$ and 4. However, the low $\chi^2_\nu$ values suggested a possible overfitting of all the redshift bins.

\subsubsection{Galaxy bias}
\label{subsubsec:subsec:bias}
From each step of the MCMC fitting posterior, we obtain a set of halo model parameters that can be used to calculate the galaxy bias according to Equation~\ref{eq:galaxy_bias}. From that, we construct the inferred posterior of the galaxy bias and therefore constrain the redshift evolution of it at $3<z<5$. For the magnitude cut sample, we observe a slight increase in galaxy bias with redshift: $b_g=3.18^{+0.14}_{-0.14}$ at $z\simeq3$, $b_g=3.58^{+0.27}_{-0.29}$ at $z\simeq4$, $b_g=4.27^{+0.25}_{-0.26}$ at $z\simeq5$. 
The inferred galaxy bias values are tightly constrained.

\begin{figure}
    \centering
    \includegraphics[width=0.48\textwidth]{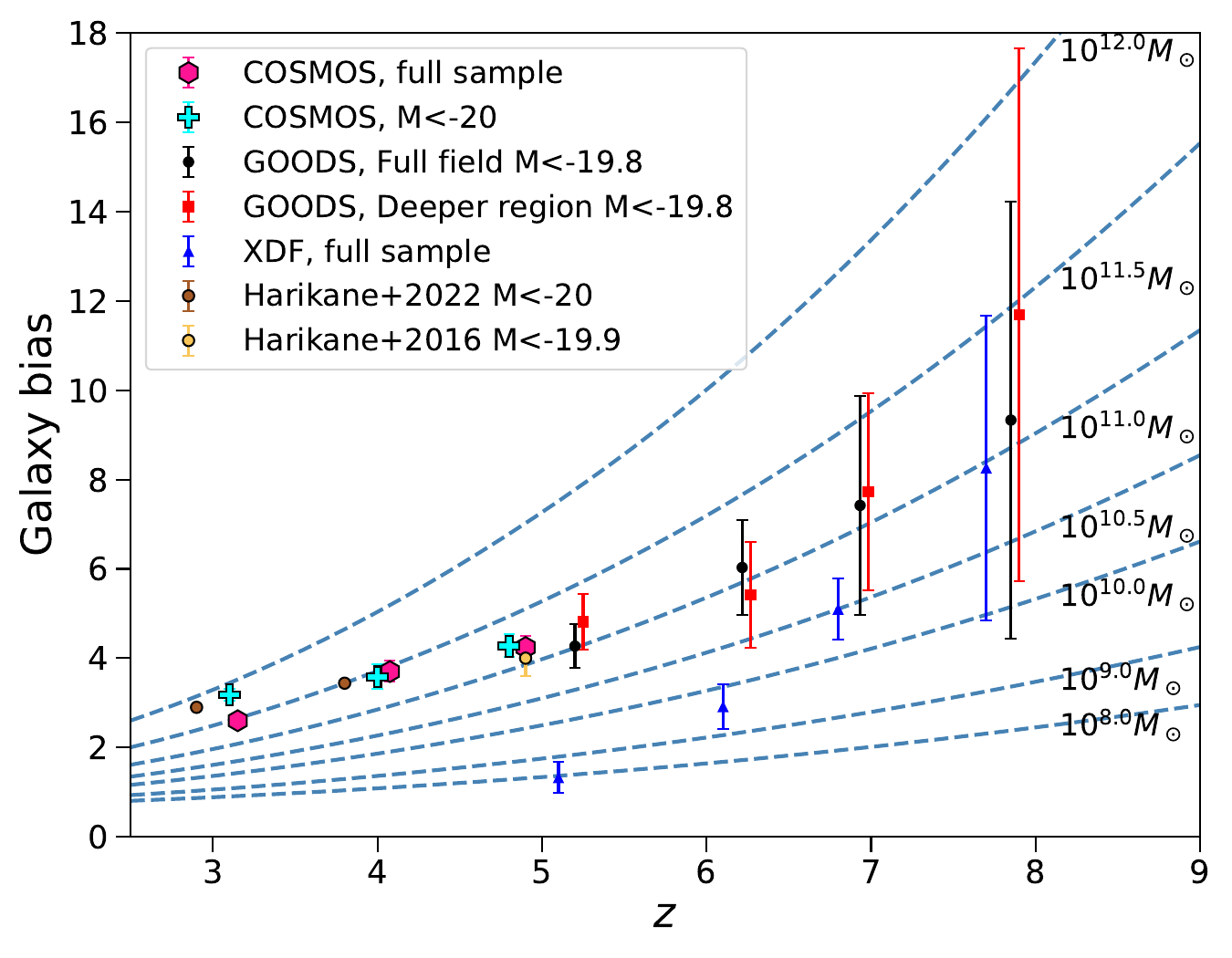}
    \caption{Galaxy bias as a function of redshifts. The hexagon- and plus-sized points are measurements from this work in the COSMOS field for the full galaxy samples and magnitude-limited samples, respectively. These are compared with previous results from GOODS and XDF fields \protect\citep{Dalmasso_2023}, and from \protect\cite{Harikane_2016,Harikane_2022}. The blue dashed lines are the expected dark matter halo bias from the \protect\cite{Tinker_2010} model for fixed halo masses. Data points are slightly shifted to the right for clarity, except for the Harikane et al. data, where no shift was necessary. Error bars represent the $1\sigma$ uncertainty from the posterior distributions obtained via MCMC fitting. For the magnitude-limited samples, the galaxy bias increases with redshift and corresponds to progressively lower halo masses as redshift increases.}
    \label{fig:bias}
\end{figure}

In Fig.~\ref{fig:bias}, we show the redshift evolution of galaxy bias from the measurements in this work. The same plot also includes results from the GOODS and XDF fields \citep{Dalmasso_2023}, reaching $z=8$, as well as results from \cite{Harikane_2016,Harikane_2022}. The bias values from this study and previous work with approximately the same magnitude cut at $3<z<5$ lie within a close range, with higher bias observed at higher redshift. In the XDF field, which shows an unusually low bias at $z\sim5$, this is because of the very low clustering amplitude measured. Note that we use a different method to calculate the galaxy bias than these studies. \cite{Dalmasso_2023} fits the ACF with a power-law, converting it to a real-space correlation length, and then derived the bias as the ratio of the galaxy correlation amplitude to the dark matter correlation amplitude at a fixed scale of 8 Mpc/h. \cite{Harikane_2016,Harikane_2022} calculated the effective galaxy bias from the halo model by weighting each halo mass by the abundance (described by the halo mass function), the average galaxy occupation in halo and the halo bias of that mass. The result is then integrated over all halo masses and normalised by total galaxy number density. 

We also present the evolution of halo bias for fixed halo masses with redshift in Fig.~\ref{fig:bias}. Comparing the measured galaxy bias to the halo biases at different halo masses, we can infer the effective halo mass of the galaxy sample. We can see that for the magnitude cut sample, the bias at $z\simeq3$ corresponds to a halo mass of approximately $10^{12} M_\odot$, at $z\simeq4$ to approximately $10^{11.5} M_\odot$, and at $z\simeq 5$ to approximately $10^{11} M_\odot$. 
This trend is consistent with the trend we see in the $M_{\rm min}$ across this redshift range.
The bias values correspond to progressively lower halo masses with increasing redshift. 

The semi-analytic model GALFORM \citep{Mitchell_2016} predicts weak evolution in the SHMR over $z=0-4$. Assuming $M_{\mathrm{UV}}$ and stellar mass are correlated at these redshifts, as shown in e.g. \cite{Speagle_2014}, the model from \textsc{UniverseMachine} simulation \citep{Behroozi_2019} similarly suggests that at fixed stellar mass, the corresponding halo mass is slightly lower at higher redshifts. At our inferred halo mass range ($10^{11}-10^{12} M_\odot$), COSMOS2020 \citep{Shuntov_2022} and the COSMOS-Web survey \citep{Paquereau_2025}, both of which compare directly to simulations such as IllustrisTNG (e.g. \citealt{Springel_2018}) finds only modest evolution in the SMHR from $z = 3-5$. In addition, models of halo star formation history (e.g. \citealt{Mason_2015}) suggest that for a fixed observed galaxy luminosity threshold, the halo mass decreases with increasing redshift. The trend seen in this study is therefore not unexpected given these previous results, however further analysis with larger samples of galaxies with robust stellar masses are needed to be able to constrain simulation predictions.

\subsubsection{Duty cycle}
\label{subsubsec:subsec:duty_cycle}

\begin{figure}
    \centering
    \includegraphics[width=0.49\textwidth]{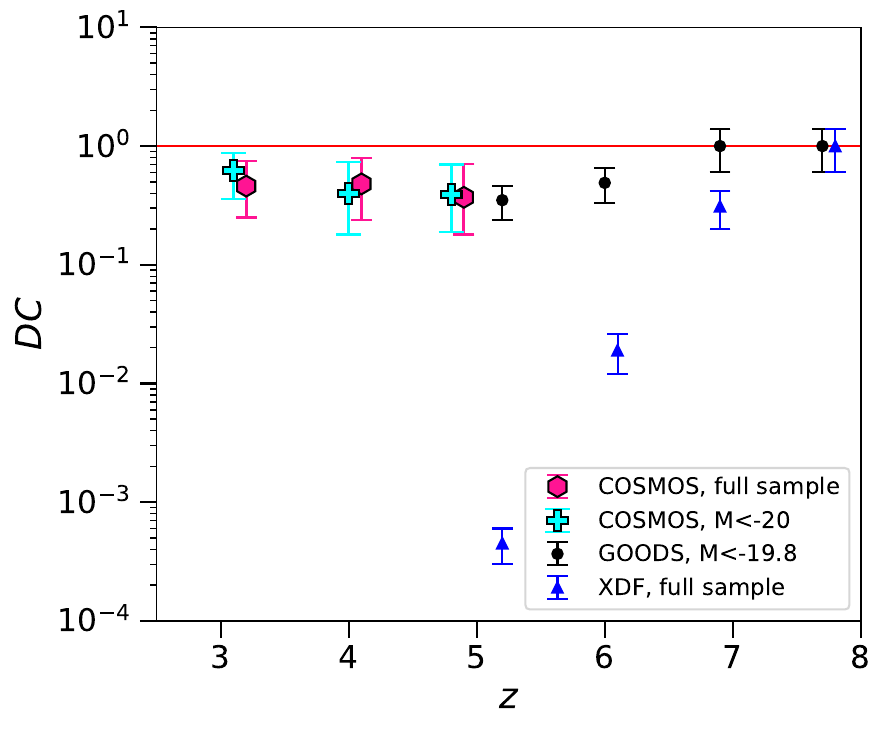}
    \caption{Duty cycle as a function of redshift. The hexagon- and plus-sized points are measurements from this work using the COSMOS dataset. The black circles and blue triangles are results from \protect\cite{Dalmasso_2023}. Some points are slightly shifted to the right for clarity. The red horizontal line represent the upper limit of 1.0. Error bars represent the $1\sigma$ uncertainty from the posterior distributions obtained via MCMC fitting.}
    \label{fig:duty_cycle}
\end{figure}

Duty cycle represents the fraction of galaxies that are bright enough to exceed a given observational threshold and hence be included in our sample. As discussed in Section~\ref{subsec:mcmc_fitting_testing}, the parameter $f_c$ can be interpreted as the duty cycle following the re-parameterisation of the satellite HOD. We obtain $f_c$ values for the $M_{\rm{UV}}$-limited samples of $0.62^{+0.25}_{-0.26}$ at $z\simeq3$, $0.40^{+0.34}_{-0.22}$ at $z\simeq4$, and $0.39^{+0.31}_{-0.20}$ at $z\simeq5$. The constraints are relatively loose however. For comparison, in Fig.~\ref{fig:duty_cycle}, we show the measurements of duty cycle from our work as well as \cite{Dalmasso_2023}, which has a similar magnitude cut.

For the magnitude-cut sample, there appears to be a very mild downward trend in $f_c$ with increasing redshift, although this trend is not significant considering the large error bar -- we do not observe a strong dependency of the duty cycle on the redshift. These values are broadly consistent with the GOODS magnitude-limited measurements at $z\sim5-7$,
which do not show a noticeable trend either. On the other hand, the measurement from the XDF field show a clear increase with redshift, although we note that the ACF in the XDF (and the higher redshift bins in GOODS) in \citet{Dalmasso_2024} are weakly detected. 

It is also important to note that \cite{Dalmasso_2023} uses an abundance matching technique to calculate the duty cycle, which is different from our method in several ways.
First, \cite{Dalmasso_2023} matches the observed galaxy number density to the number density of haloes, ignoring the contribution from satellite galaxies. Second, the number density of the halo is very sensitive to the assumed minimum halo mass. The minimum halo mass in our work is inferred from the halo model fitting, which gives a proper posterior distribution. Finally, we impose a loose abundance matching criterion as discussed in Section~\ref{subsec:mcmc_method}, where \cite{Dalmasso_2023} matches the number density exactly.

Our results of the duty cycle do not show significant evolution, and the measurements from $z=5$ to $z=4$ are consistent with the findings of \cite{Lee_2009}, where the duty cycle was also found to remain broadly constant. The value of the duty cycle indicates the fraction of haloes that host an observable LBG; a low duty cycle can suggest a short-lived star formation. The star formation timescale (or the duration of star formation) $\tau_{\mathrm{SF}}$ can be roughly estimated by relating it to the duty cycle, assuming that $\tau_{\rm{SF}} \approx DC \times \Delta t_{\mathrm{survey}}$, where $\Delta t_{\mathrm{survey}}$ is the cosmic time span probed by a given survey redshift interval \citep{Lee_2009}. For the magnitude-cut sample, we estimate the star formation timescale $\tau_{\mathrm{SF}}$ to be $\sim0.35\,\mathrm{Gyr}$ at $z=2.75-3.5$ (with a survey time span $\Delta t_{\mathrm{survey}} = 0.56\,\mathrm{Gyr}$), $\sim0.19\,\mathrm{Gyr}$ at $z=3.5-4.5$ ($\Delta t_{\mathrm{survey}} = 0.47\,\mathrm{Gyr}$), and $\sim0.09\,\mathrm{Gyr}$ at $z=4.5-5.2$ ($\Delta t_{\mathrm{survey}} = 0.22\,\mathrm{Gyr}$). The shorter timescale at higher redshift suggests more intense and rapid star-forming activity in early galaxies. \cite{Lines_2025} finds LBGs at $z\sim4-5$ have ages of $\simeq100 \,\mathrm{Myr}$, which is consistent with our inferred star formation timescale at this redshift.

\section{Discussion and conclusions}
\label{sec:discussion}

In this work we measured the angular two-point correlation function for high-redshift Lyman-break galaxies ($3 < z < 5$) selected from optical and NIR data in the COSMOS field and applied a halo model to infer the properties of their host dark matter haloes. Our analysis contributes to the currently limited number of high-redshift clustering studies, and it offers a preview of what galaxy clustering properties we might expect in future surveys like LSST, given the comparable depth. While our current measurements provide tight constraints on small (1-halo) scales, LSST’s significantly larger survey area will reduce sample variance and enable finer redshift and luminosity binning (e.g. \citealt{Cochrane_2017,Cochrane_2018}), as well as clustering studies of rarer populations like AGNs. We also investigated how varying different HOD parameters affects the shape of the model two-point function, and derived the predicted galaxy bias from our model.

We use the Landy-Szalay estimator for our correlation function measurements. At angular separations from $\theta \approx 1$ to 300~arcsec, the SNRs are high enough to yield information on clustering. For comparison and interpretive purposes, we computed the angular correlation function in each redshift bin for both the full and magnitude-limited ($M_{\rm{UV}}<-20.0$) subsample. Overall, the clustering amplitude increases with redshift, primarily because brighter galaxies are being selected. Luminous galaxies are typically hosted by more massive haloes, which are associated with a larger galaxy bias. Consequently, these galaxies exhibit stronger clustering. Note that at higher redshifts, dark matter is more clustered as well, which contributes to the change in the amplitude of the angular correlation function.

The steeper power-law behaviour observed at the smallest angular separations ($<3$~arcsec) is likely driven by physical processes not captured by the HOD formalism, such as merger-induced star formation. To avoid biasing the fit, we excluded these small-scale data points from the halo model fitting, as the enhanced UV luminosity and star formation in close interacting systems can affect the observed number of LBGs. In Appendix~\ref{appendixpairs} and Fig.~\ref{fig:pairimages} we highlight how some of these close sources are likely star-forming clumps in the same galaxy or interacting galaxies.

In the halo model fitting analysis, we tested various combinations of the five free parameters $\log_{10} M_{\rm min}$, $\log_{10} M_1$, $\alpha$, $f_c$, $\beta_g$, and found that allowing all five parameters to vary provided the best fit. We thus set all five parameters as free in the fitting. We obtained good fits across the full range of scales, including the transition between the 1-halo and 2-halo terms for the $z\sim 3$ and 4 samples, whereas previous studies, such as \cite{Harikane_2022}, excluded data points at these scales in their fits. At around $10-20$~arcsec, the 1-halo term, which indicates pairs of galaxies are more likely to reside within the same halo, was seen to decline with redshift. At these scales, clustering shifts from being dominated by galaxy pairs within a single halo to galaxy pairs residing in separate haloes. Beyond this scale, the 2-halo term dominates, and we calculated the galaxy bias from this term. 

We used the linear matter power spectrum as the input for our HOD calculations, which is as intended by the theory. As a consistency check,  we tested the impact of using the non-linear matter power spectrum as the input instead following \cite{Jose_2017}. Specifically, we used the `halofit' model \citep{Takahashi_2012} implemented in $\textsc{pyccl}$, and subsequently the smaller scales have more contribution from the 2-halo term when taken account of the nonlinearity. \cite{Harikane_2022} exclude measurements at the quasi-linear scales, since they are potentially affected by the non-linear halo bias effect. Nevertheless, we achieve good fits using the linear matter power spectrum, suggesting that using the non-linear spectrum may not be necessary.

When including the data points at pairing scales $<20$~kpc, the posterior of the $\beta_g$ parameter hits the physical prior $\beta_g>0$, and gives very small values of $\beta_g \sim0.01$. If we keep the intended interpretation of $\beta_g$ as the ratio of satellite galaxy location to the halo scale radius, this would indicate that most satellite galaxies are located extremely close (approximately $1\%$ of the halo scale radius) to the central galaxy position. This is physically unrealistic and implies that the measured clustering signal cannot be fully described by the simple 1-halo term alone. This is further justification for removing the small scale points ($\theta<3^{\prime\prime}$) in the HOD fitting.

In the fitting, one of the free parameters, $\alpha$, can compensate for changes in $\log_{10}M_1$. Many previous studies avoid setting $\alpha$ as a free parameter; however, fixing it to a known value can introduce bias in $\log_{10}M_1$ if the chosen value of $\alpha$ is inaccurate. In our analysis, we found that the value of 1 lies outside the 3$\sigma$ range of the posterior distribution. The parameter $f_c$ can be treated as the duty cycle (the probability that we observe a given central LBG or not, depending on whether it is in an actively star-forming vs. undetectable phase) by simply modifying the measured $M_1$. Note that the uncertainties from the fitting are still quite large; we would expect future larger-scale surveys with substantially more galaxies to help better constrain the dark matter halo properties.

In the magnitude-limited sample, we observe a mild increase in galaxy bias across redshifts $z\simeq3-5$: $b_g=3.18^{+0.14}_{-0.14}$ at $z\simeq3$, $b_g=3.58^{+0.27}_{-0.29}$ at $z\simeq4$, $b_g=4.27^{+0.25}_{-0.26}$ at $z\simeq5$. This trend corresponds to a decrease in the effective halo mass at higher redshift for a fixed UV luminosity threshold (for the inferred $10^{11}-10^{12} M_\odot$ haloes). We also constrained the observed central galaxy fraction $f_c$ which we interpret as the galaxy duty cycle. This interpretation slightly modifies the inferred value of $M_1$. While the inferred $f_c$ values are loosely constrained, they appear to show no evolution or a weak decline with redshift: $\sim0.62^{+0.25}_{-0.26}$ at $z\simeq3$, $\sim0.40^{+0.34}_{-0.23}$ at $z\simeq4$,  $\sim0.39^{+0.31}_{-0.20}$ at $z\simeq5$. These results are in broad agreement with previous measurements by \citet{Lee_2009}, who also found loose constraints and no significant evolution from $z=5$ to $z=4$. The rough estimates of star formation timescale inferred from the duty cycle are $\sim0.35\,\mathrm{Gyr}$, $\sim0.19\,\mathrm{Gyr}$, and $\sim0.09\,\mathrm{Gyr}$ at $z\sim3,4,5$, respectively, consistent with the observed ages of typical Lyman-break galaxies at these redshifts (e.g.~\citealp{Lines_2025}).

Overall, our correlation function measurements of high-redshift LBGs in the COSMOS field enabled a robust halo model fit, particularly capturing the clustering well at intermediate scales where the 1-halo and 2-halo terms intersect at $z\simeq3$ and 4. This modeling provides constraints on the properties of their host dark matter haloes, and gives a preview of future studies of galaxy clustering at these redshift, as well as informing modelling of the high-$z$ galaxy population's clustering in forthcoming surveys with similar depth but much larger area, such as LSST.

\section*{Acknowledgements}
We are grateful to S.~Cunnington, C.~D.~Leonard, S.~Kay, and A.~Nicola for useful discussions.
This result is part of a project that has received funding from the European Research Council (ERC) under the European Union's Horizon 2020 research and innovation programme (grant agreement no. 948764; PB). 
IY acknowledges support from the China Scholarship Council (grant no. 202208060318).
RB acknowledges support from an STFC Ernest Rutherford Fellowship [grant no. ST/T003596/1]. RKC is grateful for support from the Leverhulme Trust via the Leverhulme Early Career Fellowship. NJA  acknowledges support from the ERC Advanced Investigator Grant EPOCHS (788113). MJJ acknowledges the support of a UKRI Frontiers Research Grant [EP/X026639/1], which was selected by the European Research Council, the STFC consolidated grant [ST/W000903/1], and support from the Oxford Hintze Centre for Astrophysical Surveys which is funded through generous support from the Hintze Family Charitable Foundation. 

\balance

\section*{Data Availability}
The image data are publicly available from the respective survey archives. The computer code used in this study is available on request.



\bibliographystyle{mnras}
\bibliography{clustering_highz} 




\appendix

\section{Examples of galaxy close pairs}\label{appendixpairs}

\begin{figure*}
    \centering
    \includegraphics[trim = 0.5cm 1cm 0.5cm 1cm, width=\linewidth]{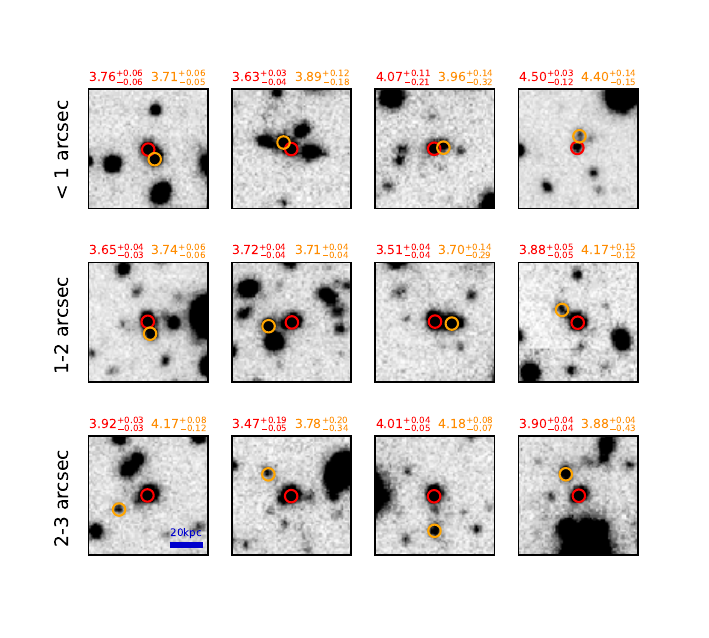}
    \caption{Postage-stamp images in the HSC $i$-band of 12 galaxies in the $z \simeq 4$ sample we use for ACF measurements (from~\citealt{adams2023total}) that show close separations to another source in the sample.
    From top to bottom the rows correspond to separations of $< 1$, $1$--$2$ and $2$--$3$arcsec respectively, and the central galaxies are taken from the brightest magnitude bin ($M_{\rm{UV}} \simeq -22$).
    The photometric redshift of the central source (red) and the companion (orange) are shown along the top of each stamp.
    The stamps are 10 arcsec on a side with North to the top and East to the left.
    They have been scaled from [-2, 8] $\sigma$, where $\sigma$ is the local pixel RMS.  
    }
    \label{fig:pairimages}
\end{figure*}

In Fig.~\ref{fig:pairimages} we show image cut-outs centered on a selection of the brightest galaxies within the $z \simeq 4$ sample.
These sources were identified to have close separations to another galaxy within the same redshift bin.
We find an excess in the ACF on small scales ($< 3$ arcsec) in comparison to that expected from the 1-halo term fitted to the data at larger scales.
Typically in ACF fitting the smallest scales are excluded from the fitting due to the uncertain impact of astrophysical effects such as merger-induced star-formation on scales of several tens of physical kpc.
We visually inspected the close pairs in our sample to investigate this further.
At the smallest separations ($< 1$ arcsec) we see that these sources are all likely to be close mergering pairs or clumps within the same galaxy.
The photometric redshifts for these close pairs are all consistent within $1$--$3\sigma$.
Moving to wider separations, at $1$--$3$ arcsec, we further see evidence for physical association, with flux extending between the two sources that again have consistent redshifts within the errors.
There is also some evidence visually for tidal tails, which would support a merger origin for the pairs.
Mergers have been shown to enhance the SFR of galaxies at separations up to $\sim 20\,{\rm kpc}$~\citep{Scudder_2012, Duan_2024}.
As our sample selection is based on the rest-frame UV magnitude (a proxy for recent star formation) it is possible that our sample is biased towards merging sources, causing the observed excess in the ACF on scales of $<3$ arcsec.

\section{Two-dimensional posterior distribution of halo fitting}\label{corner_plots}

\begin{figure*}
    \centering
    \includegraphics[width=0.85\textwidth]{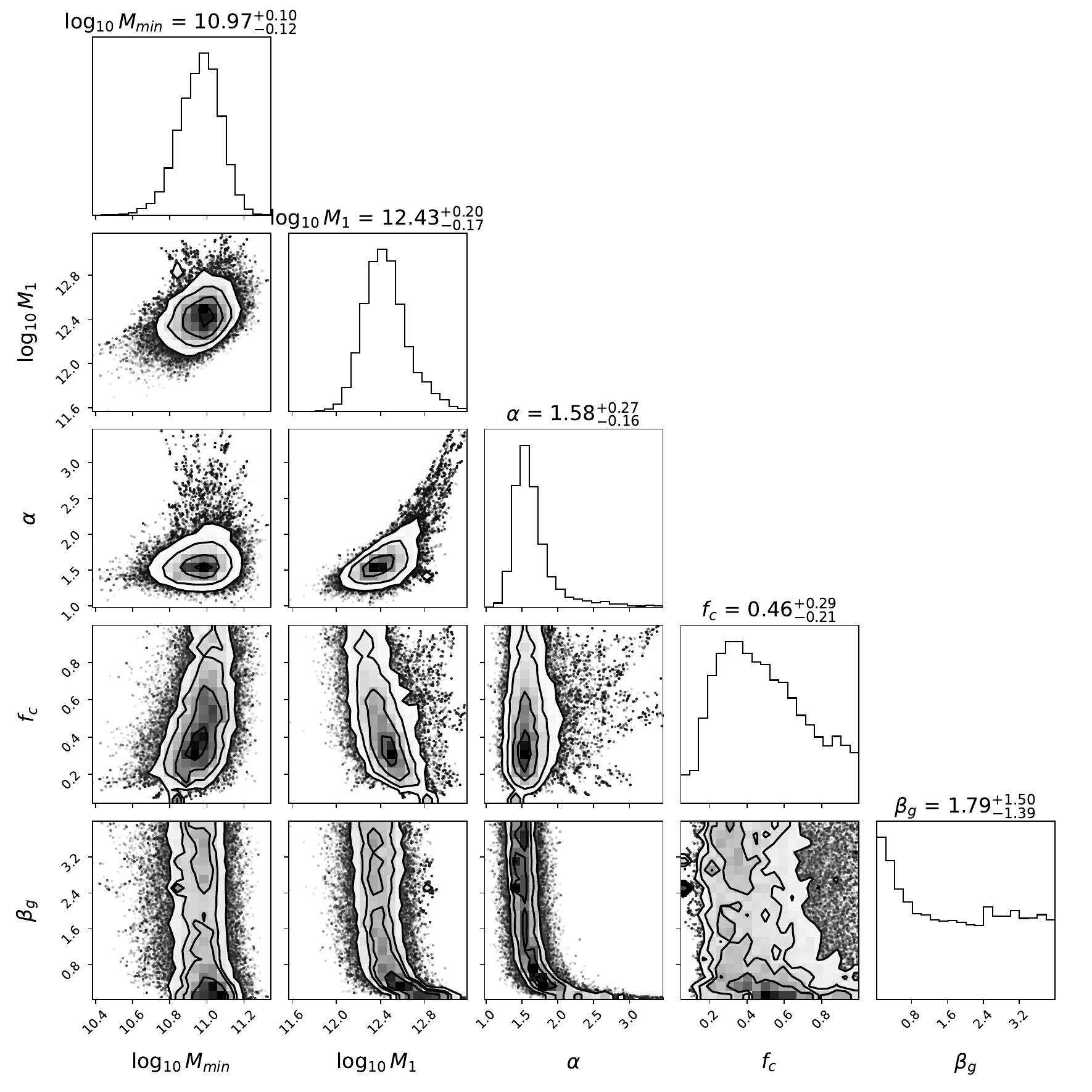}
    \caption{Corner plot showing marginalised posterior distributions and parameter correlations from MCMC halo model fitting at $z\simeq3$ for the full sample. The parameters include $\log_{10}M_{\text{min}}$, $\log_{10}M_1$, $\alpha$, $f_c$, $\beta_g$. The contours represent 68$\%$ and 95$\%$ confidence intervals for the parameter pairs. Marginalised 1D distributions with median values and 68$\%$ confidence intervals for each parameter are shown in the diagonal panels.}
    \label{fig:contourplot_z3}
\end{figure*}

\begin{figure*}
    \centering
    \includegraphics[width=0.85\textwidth]{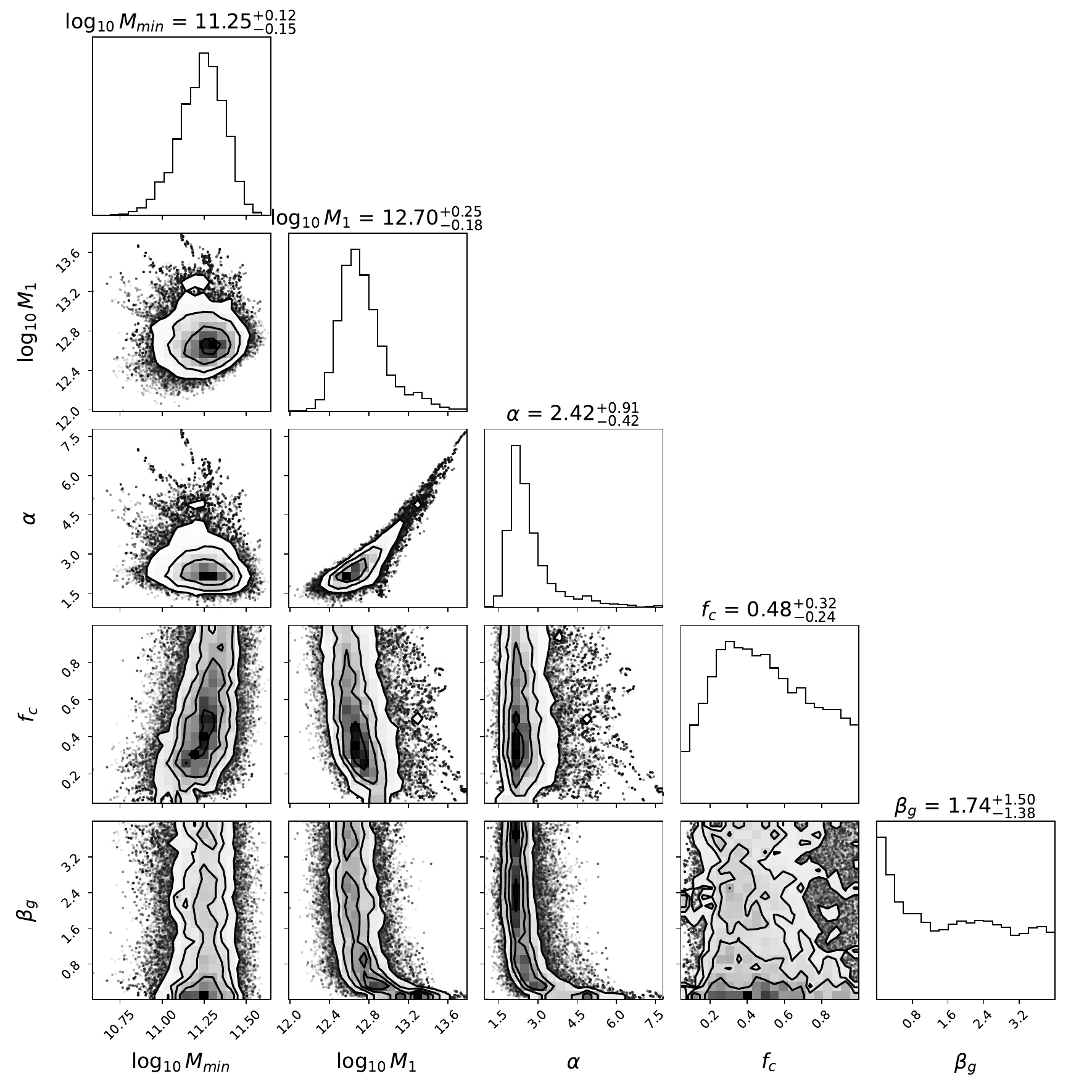}
    \caption{Corner plot showing marginalised posterior distributions and parameter correlations from MCMC halo model fitting at $z\simeq4$ for the full sample. The parameters include $\log_{10}M_{\text{min}}$, $\log_{10}M_1$, $\alpha$, $f_c$, $\beta_g$. The contours represent 68$\%$ and 95$\%$ confidence intervals for the parameter pairs. Marginalised 1D distributions with median values and 68$\%$ confidence intervals for each parameter are shown in the diagonal panels.}
    \label{fig:contourplot_z4}
\end{figure*}

\begin{figure*}
    \centering
    \includegraphics[width=0.85\textwidth]{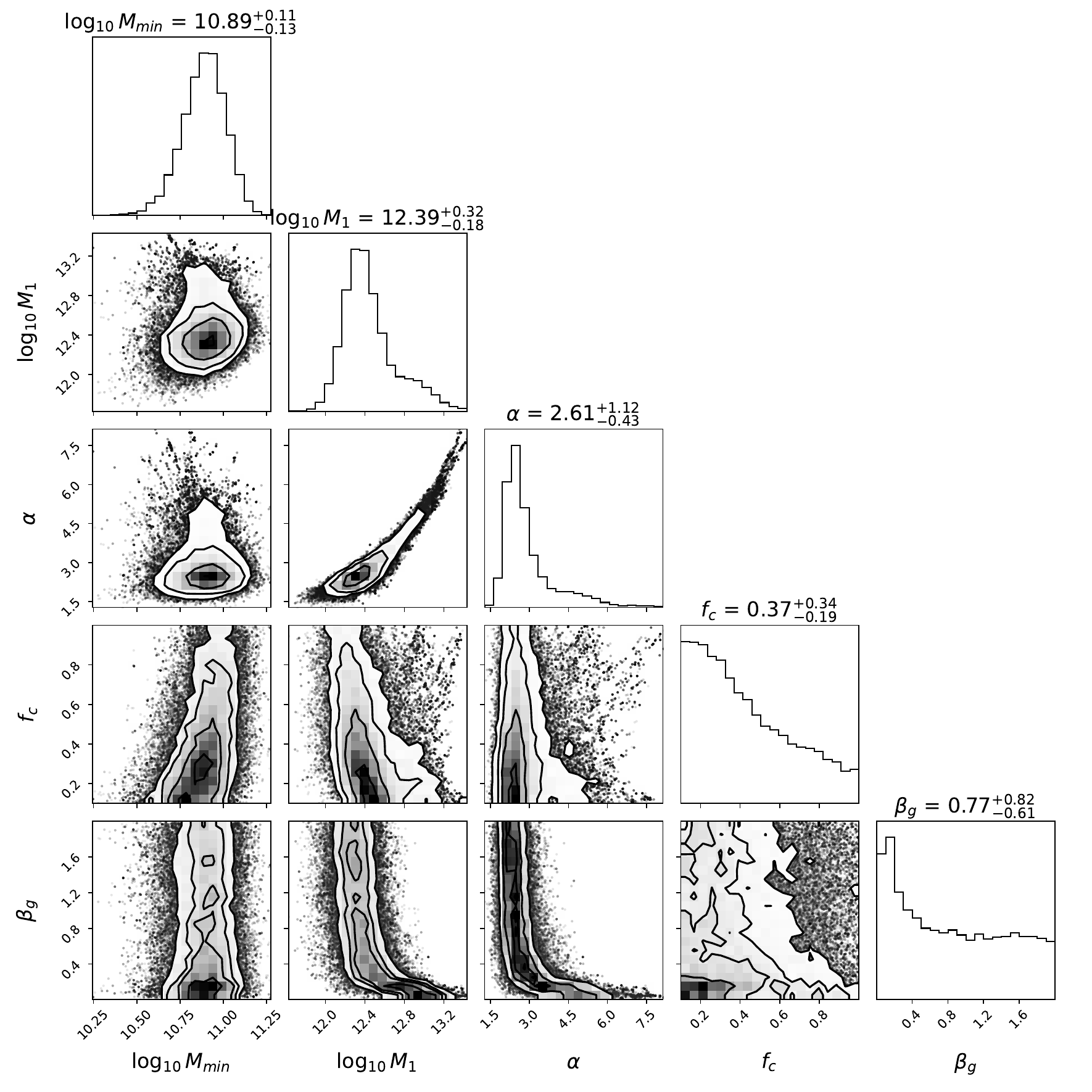}
    \caption{Corner plot showing marginalised posterior distributions and parameter correlations from MCMC halo model fitting at $z\simeq5$ for the full sample. The parameters include $\log_{10}M_{\text{min}}$, $\log_{10}M_1$, $\alpha$, $f_c$, $\beta_g$. The contours represent 68$\%$ and 95$\%$ confidence intervals for the parameter pairs. Marginalised 1D distributions with median values and 68$\%$ confidence intervals for each parameter are shown in the diagonal panels.}
    \label{fig:contourplot_z5}
\end{figure*}

Figs.~\ref{fig:contourplot_z3}, \ref{fig:contourplot_z4}, and \ref{fig:contourplot_z5} illustrate the two-dimensional posterior distributions for the $z\simeq3$, $z\simeq4$, and $z\simeq5$ full sample fits, respectively. 

\bsp	
\label{lastpage}
\end{document}